\documentclass[aps,pra,10pt,amsmath,notitlepage,nofootinbib,twocolumn,superscriptaddress]{revtex4-2}
\pdfoutput=1
\usepackage[utf8]{inputenc}
\usepackage[english]{babel}
\usepackage[T1]{fontenc}
\usepackage{microtype}

\usepackage{amsfonts}
\usepackage{amsmath}
\usepackage{amssymb}
\usepackage{graphicx}
\usepackage[dvipsnames]{xcolor}

\usepackage[breaklinks=true,colorlinks=true,linkcolor=teal,urlcolor=teal,citecolor=teal]{hyperref}
\usepackage{orcidlink}

\let\originalleft\left
\let\originalright\right
\renewcommand{\left}{\mathopen{}\mathclose\bgroup\originalleft}
\renewcommand{\right}{\aftergroup\egroup\originalright}

\DeclareMathOperator{\csch}{csch}
\DeclareMathOperator{\Tr}{Tr}
\DeclareMathOperator{\tr}{tr}

\newcommand{\I}{\mathrm{i}}

\let\Re\relax
\let\Im\relax
\DeclareMathOperator{\Re}{Re}
\DeclareMathOperator{\Im}{Im}

\renewcommand{\vec}{\boldsymbol}

\newcommand{\Dsuper}{\mathcal{D}_{\!\rho_{\theta}}\!}

\newcommand{\CSfull}{C_{\theta}^{\mathrm{S}}(W)}
\newcommand{\CRfull}{C_{\theta}^{\mathrm{R}}(W)}
\newcommand{\CHfull}{C_{\theta }^{\mathrm{H}}(W)}
\newcommand{\CUfull}{\overline{C}_{\theta }^{\mathrm{H}}(W)}
\newcommand{\mS}{\mathrm{S}}
\newcommand{\mR}{\mathrm{R}}

\begin{document}

\title{Multiparameter quantum estimation with Gaussian states: efficiently evaluating Holevo, RLD and SLD Cram\'er-Rao bounds}

\author{Shoukang Chang\,\orcidlink{0000-0002-1824-1668}}
\email{changshoukang@stu.xjtu.edu.cn}
\affiliation{MOE Key Laboratory for Nonequilibrium Synthesis and Modulation of Condensed Matter, Shaanxi Province Key Laboratory of Quantum Information and
Quantum Optoelectronic Devices, School of Physics, Xi’an Jiaotong University, Xi’an 710049, People’s Republic of China}
\affiliation{Department of Physics ``A. Pontremoli", Universit\`a degli Studi di Milano, I-20133 Milano, Italy}%

\author{Marco G. Genoni\,\orcidlink{0000-0001-7270-4742}}
\email{marco.genoni@unimi.it}
\affiliation{Department of Physics ``A. Pontremoli", Universit\`a degli Studi di Milano, I-20133 Milano, Italy}%

\author{Francesco Albarelli\,\orcidlink{0000-0001-5775-168X}}
\email{francesco.albarelli@gmail.com}
\affiliation{Scuola Normale Superiore, I-56126 Pisa, Italy}

\begin{abstract}
Multiparameter quantum estimation theory is crucial for many applications involving infinite-dimensional Gaussian quantum systems, since they can describe many physical platforms, e.g., quantum optical and optomechanical systems and atomic ensembles.
In the multiparameter setting, the most fundamental estimation error (quantified by the trace of the estimator covariance matrix) is given by the Holevo Cram\'{e}r-Rao bound (HCRB), which takes into account the asymptotic detrimental impact of measurement incompatibility on the simultaneous estimation of parameters encoded in a quantum state.
However, the difficulty of evaluating the HCRB for infinite-dimensional systems weakens the practicality of applying this tool in realistic scenarios.
In this paper, we introduce an efficient numerical method to evaluate the HCRB for general Gaussian states, by solving a semidefinite program involving only the covariance matrix and first moment vector and their parametric derivatives.
This approach follows similar techniques developed for finite-dimensional systems, and hinges on a phase-space evaluation of inner products between observables that are at most quadratic in the canonical bosonic operators.
From this vantage point, we can also understand symmetric and right logarithmic derivative scalar Cramér-Rao bounds under the same common framework, showing how they can similarly be evaluated as semidefinite programs.
To exemplify the relevance and applicability of this methodology, we consider two paradigmatic applications, where the parameter dependence appears both in the first moments and in the covariance matrix of Gaussian states: estimation of phase and loss, and estimation of squeezing and displacement.
\end{abstract}

\maketitle

% \section{Introduction}

The goal of quantum parameter estimation theory is to assess the fundamental precision limits on measuring unknown parameters characterizing quantum systems, and to find practical strategies to attain them.
This endeavor has technological applications, such as quantum-enhanced metrology and sensing~\cite{Giovannetti2011, Demkowicz-Dobrzanski2015a,Degen2016,Pezze2018,Pirandola2018,Polino2020,Barbieri2022,Liu2024s}, but is also intimately connected with deep fundamental aspects of quantum physics, such as the geometry of quantum states~\cite{bengtsson2017geometry} and incompatibility of observables.

Single-parameter quantum estimation has been studied intensely and has been applied to devise quantum-enhanced strategies for measuring various physical parameters.
When many experimental shots are available, the attainable precision limit is the quantum Cramér-Rao bound (CRB), originally introduced by Helstrom~\cite{helstrom1976quantum,Holevo2011b,Nagaoka1989,Braunstein1994,Paris2009}.
In many situations, however, one needs to estimate multiple unknown parameters simultaneously~\cite{Szczykulska2016,Albarelli2019c,Demkowicz-Dobrzanski2020}, e.g. orthogonal displacements~\cite{Yuen1973,Genoni2013b,Bradshaw2017,Park2022,Hanamura2023,Zhou2025a,Frigerio2025} relevant in waveform estimation~\cite{Gardner2024}, multiple  phases~\cite{Humphreys2013,Pezze2017,Gorecki2022a,Gebhart2021,Chesi2023,Barbieri2024a}, phase and noise parameters~\cite{Pinel2013,Crowley2014,Altorio2015a,Roccia2018,Szczykulska2017,Roccia2018,Jayakumar2024}, relative positions of discrete optical sources~\cite{Napoli2018,Bisketzi2019,Fiderer2021} or moments of extended sources~\cite{Tsang2018,Zhou2019}, range and speed of moving targets~\cite{Huang2021,Reichert2024}, vector fields~\cite{Vaneph2012,Baumgratz2015,Hou2020,Gorecki2022b,Kaubruegger2023}, and more.

From a theoretical perspective, precision limits for multiparameter estimation were first obtained by considering the formal similarity between quantum theory and probability theory, obtaining a ``quantization'' of the classical CRB.
This procedure is not unique and leads to different Cram\'{e}r-Rao bounds~\cite{Hayashi2017c}---
the most well-known being the symmetric logarithmic derivative Cram\'{e}r-Rao bound (SLD-CRB) and the right logarithmic derivative Cram\'{e}r-Rao bound (RLD-CRB)~\cite{Yuen1973,Belavkin1976,Holevo2011b}.
These have been widely used due to their relatively easy calculation; nevertheless, neither of them is generally tight~\cite{Suzuki2018,Ragy2016}.
A tighter precision limit is the Holevo Cram\'{e}r-Rao bound (HCRB), which is uniquely quantum and not obtained by the formal analogy between quantum states and probabilities.
Crucially, the HCRB also has a fundamental operational meaning: it can be attained by collective measurements on asymptotically many identical copies of the quantum state~\cite{Yang2018a,Demkowicz-Dobrzanski2020}.
Moreover, in some cases it is also attainable using single-copy measurements: for pure states~\cite{Matsumoto2002} and for displacement estimation with Gaussian states~\cite{Holevo2011b}.

Despite its critical importance, the HCRB is often avoided in favor of bounds that are simpler to evaluate, as it lacks a general closed-form expression and is defined via a minimization.
Luckily, this is a convex optimization problem and can be expressed as a semidefinite program (SDP) for finite-dimensional systems~\cite{Albarelli2019}, which makes a numerical evaluation relatively straightforward.
For continuous variable (CV) quantum systems, the infinite dimension of the Hilbert spaces generally does not allow this simpler formulation.
Truncating the Hilbert space dimension to obtain approximations is in principle possible, but the procedure quickly becomes unwieldy.
Restricting to the more manageable class of Gaussian states~\cite{serafini2017quantum}, a semidefinite program to evaluate the HCRB for joint displacement estimation is known~\cite{Bradshaw2017}, but not for more general parameters encoded also in the covariance matrix.

We fill this gap by presenting a general SDP to evaluate the HCRB for infinite-dimensional Gaussian quantum systems. 
This allows us to study the simultaneous estimation of multiple incompatible parameters generically encoded in Gaussian quantum states of an arbitrary, but finite, number of bosonic modes.
Essentially, this is achieved by following a geometric approach to quantum estimation theory, based on the introduction of Hilbert spaces for operators, with suitable inner products defined in terms of the quantum state~\cite{Holevo2011b,Tsang2019}.
As illustrative applications of these tools, we investigate two paradigmatic models for which the HCRB is tighter than both the SLD-CRB and the RLD-CRB: the joint estimation of phase and loss, which has been mainly studied in the literature in terms of the SLD-CRB~\cite{Pinel2013,Crowley2014,Ragy2016,Nichols2018,Gianani2021}, and the estimation of complex displacement and squeezing for single- and two-mode Gaussian states, for which the HCRB has been recently evaluated for pure Gaussian states~\cite{Bressanini2024a}, extending previous results on the estimation of displacement parameters only~\cite{Genoni2013b,Bradshaw2017,Bradshaw2017a}.

In this paper, we abstain from introducing too many abstract mathematical concepts and notations, having the goal to provide results that are immediately accessible and useful for practitioners of quantum estimation with continuous variable systems.
Despite this effort, the derivations remain rather technical, so we provide a summary of the main results in Sec.~\ref{sec:overview}.
This section contains all the practical information needed to compute multiparameter bounds for Gaussian states, including the SLD-CRB and RLD-CRB for Gaussian states already known in the literature~\cite{Monras2013,Gao2014a,Jiang2014,Serafini2023,Nichols2018,Safranek2019}.
These bounds are presented in a unified and familiar notation, allowing also to show how they can be also obtained by solving a SDP.

This paper is arranged as follows.
In Sec.~\ref{sec:preliminaries} we review the elementary framework of multiparameter quantum estimation theory and the formalism to describe Gaussian quantum systems. 
Sec.~\ref{sec:overview} contains the aforementioned overview of the main results, while their derivation is presented in Sec.~\ref{sec:derivation}.
In Sec.~\ref{sec:examples} we demonstrate our approach with two specific applications.
Finally, Sec.~\ref{sec:conclusions} concludes the paper with some remarks and open questions.

\section{Methods}\label{sec:preliminaries}
In this section, we first introduce the main mathematical tools and definitions that allow us to derive the ultimate limits for multiparameter quantum estimation protocols, then we present the formalism needed to describe continuous-variable Gaussian quantum systems, and finally we give a short note on the vectorization procedure for finite-dimensional matrices.

\subsection{Multiparameter quantum estimation theory}

Quantum parameter estimation theory examines the ultimate precision limits of parameter estimation in quantum states and the measurements that achieve them.
Our focus is the simultaneous estimation of $p$ unknown parameters $\theta $=$(\theta _{1},...,\theta _{p})^{\! \top}$ characterizing a parametric family of quantum states $\hat{\rho}_{\theta}$, which we also call a quantum statistical model.
To estimate these unknown parameters, one performs a generic quantum measurement, mathematically described by a positive operator-valued measure (POVM).
The corresponding probability for the measurement outcome $x$ is given by Born's rule $P(x|\theta ) = \Tr(\hat{\rho}_{\theta }\hat{\Pi}_{x})$ where $\hat{\Pi}_{x}$ is the POVM element.
Based on the observed measurement outcome $x$, one can use an appropriate estimator function $\check{\theta}(x)$ to infer the parameter values.
We quantify the quality of the estimator $\check{\theta}(x)$ by the mean square error matrix (MSEM):%
\begin{equation}
\Sigma_{\theta }(\check{\theta})=\int \! dx \, P(x|\theta ) \left(\check{\theta}(x)-\theta \right) \left(\check{\theta}(x)-\theta \right)^{\! \top}. 
\end{equation}
Imposing the local unbiasedness conditions
\begin{eqnarray}
\int \! dx \, P(x|\theta )(\check{\theta}_{j}(x)-\theta _{j}) &=&0,  \notag
\\
\int \! dx \, \check{\theta}_{j}(x)\frac{\partial P(x|\theta )}{\partial \theta _{k}%
} &=&\delta _{jk}, 
\end{eqnarray}
on the estimator $\check{\theta}(x)$ one can derive the matrix Cram\'{e}r-Rao bound (CRB)\footnote{A matrix inequality $A \geq B$ means that the matrix $A-B$ is positive semidefinite} on the MSEM~\cite{cramer1946mathematical,Braunstein1994}
\begin{equation}
\Sigma _{\theta }(\check{\theta}(x))\geq F^{-1},  \label{8}
\end{equation}%
where $F$ is the classical Fisher information matrix (FIM) with matrix elements
\begin{equation}
F_{jk}=\int dxP(x|\theta ) \left( \frac{\partial \ln P(x|\theta )}{
\partial \theta_{j}} \right)\left( \frac{\partial \ln P(x|\theta ) }{\partial
\theta _{k}} \right).
\end{equation}

Since for a given parametric family of quantum states $\hat{\rho}_{\theta}$ infinitely many measurement choices are possible, one can introduce quantum precision bounds that hold for any POVM and depend only on $\hat{\rho}_{\theta}$.
A well-known quantum lower bound on the MSEM is given by the replacing the CFIM with the symmetric logarithmic derivative quantum Fisher information matrix (SLD-QFIM) introduced by Helstrom~\cite{Helstrom1967,Liu2019d}:
\begin{align}
J_{jk}^{\mS} = \Re\left( \Tr[\hat{L}_{j}^{\mS}\hat{\rho}_{\theta }\hat{L}_{k}^{\mS}] \right) = \frac{1}{2} \Tr \left[ \hat{\rho}_{\theta} \left\{ L^{\mS}_k , L^{\mS}_j \right\}  \right]\,,
\label{eq:SLD-QFIelementsDEF}
\end{align}
where the symmetric logarithmic derivative (SLD) operators $\hat{L}_{j}^{\mS}$ are defined as solutions of the equation~\cite{Helstrom1967,Braunstein1994,Paris2009}
\begin{align}
2 \frac{\partial \hat{\rho}
_{\theta }}{\partial \theta _{j}} = \{ \hat{\rho}_{\theta },\hat{L}_{j}^{\mS}\} \, ,
\label{eq:SLDoperator}
\end{align}
where curly brackets denote the anticommutator.
The quantity $\Re( \Tr[ \hat{Y} \hat{\rho}  \hat{X} ] )$ is known as the SLD inner product between the two Hermitian operators $\hat{X}$ and $\hat{Y}$.
The SLDs are Hermitian and have zero mean, i.e. $\Tr[\rho_\theta \hat{L}_{j}^{S}]=0$, since the partial derivatives $\partial \hat{\rho}_{\theta } / {\partial \theta _{j}}$ have trace zero as a consequence of the normalization of $\rho_\theta$.
The SLD-QFIM is real-valued, symmetric and positive definite.
We assume here and in the rest of the manuscript strict positivity, which excludes singular quantum statistical models, where not all parameters can be estimated independently (we will briefly comment about relaxing this assumption in the conclusions, Sec.~\ref{sec:conclusions}).
Unlike the single-parameter case, the SLD-CRB is in general not tight in multiparameter estimation, because the best measurements for different parameters may be incompatible.

Another well-known bound is obtained from the complex-valued and positive definite right logarithmic derivative (RLD) QFI matrix (RLD-QFIM)
\begin{align}
J_{jk}^{\mR}=\Tr[\hat{L}_{j}^{\mR\dagger }\hat{\rho}_{\theta }\hat{L}_{k}^{\mR}]\,,
\label{eq:RLD-QFIelementsDEF}
\end{align}
where the RLD operators $\hat{L}_{j}^{\mR}$ are defined as~\cite{Yuen1973,Belavkin1976}
\begin{align}
\frac{\partial \hat{\rho}_{\theta }}{\partial \theta
_{j}} = \hat{\rho}_{\theta }\hat{L}_{j}^{\mathrm{R}}\,,
\label{eq:RLDoperator}
\end{align}
these are in general non-Hermitian operators, but still zero mean, $\Tr[\hat{\rho}_\theta \hat{L}_{j}^{\mR}]=0$.
The quantity $\Tr[ \hat{Y}^\dag \hat{\rho} \hat{X} ]$ is known as the RLD inner product between the operators $\hat{X}$ and $\hat{Y}$ (in this case not necessarily Hermitian).

It is customary to quantify the overall precision with the scalar quantity\footnote{Notice that we denote with $\tr[\bullet]$ the trace of finite-dimensional $p{\times}p$ matrices, while we reserve $\Tr[\bullet]$ for the trace of Hilbert space operators.} $\tr \left[ W \Sigma_{\theta }(\check{\theta}) \right]$, where the weight matrix $W>0$ is real and symmetric.
This is a well-defined figure of merit to assess the precision of multiparameter estimation strategies, and to compare different matrix bounds.
Indeed, from the matrix inequalities on the MSEM $\Sigma_{\theta}(\check{\theta})$ in terms of SLD-QFIM and RLD-QFIM, we obtain the following scalar SLD-CRB and RLD-CRB~\cite{Demkowicz-Dobrzanski2020,Albarelli2019c}
\begin{eqnarray}
\tr \left[ W \Sigma_{\theta }(\check{\theta}) \right] \geq \CSfull &=& \tr \left[ W(J^{S})^{-1} \right],
\label{eq:SLDCRB} \\
\tr \left[ W \Sigma_{\theta }(\check{\theta}) \right] \geq \CRfull &=& \text{tr}\left[ W\Re
\left[ (J^{\mR})^{-1} \right]   \right] \label{eq:RLDCRB}   \\
&&\,\,\, + \left \Vert \sqrt{W}\Im\left[ (J^{\mR})^{-1} \right]% 
\sqrt{W}\right \Vert _{1}, \notag
\end{eqnarray}%
with $\left \Vert A\right \Vert _{1} = \tr(\sqrt{A^{\dagger}A})$ denoting
the trace norm; the real and imaginary parts of matrices are taken element-wise.
Neither of the bounds is always tight

Having a scalar figure of merit, it is natural to introduce the so-called most informative bound, i.e. the classical CRB optimized over POVMs~\cite{Nagaoka1989}:
\begin{equation}
\tr \left[ W \Sigma_{\theta }(\check{\theta}) \right] \geq C^{\mathrm{MI}}_{\theta}(W) \equiv \min_{\text{POVM}}[\text{tr(}WF^{-1}%
\text{)}].
\label{eq:CMI}
\end{equation}
This bound is achievable with many repetitions, yet no simple recipe for its evaluation is known, even though it has been recently recast as a conic optimization for finite-dimensional systems~\cite{Hayashi2023a}.

The bound $C^{\mathrm{MI}}$ holds at the single copy level and does not take into account the possibility of measuring multiple copies at the same time.
The attainable precision for collective measurement over asymptotically many identical copies of the state $\hat{\rho}_{\theta }^{\otimes n}$ is given by the Holevo Cram\'{e}r-Rao bound (HCRB)~\cite{Holevo2011b,Holevo1973,Nagaoka1989,Demkowicz-Dobrzanski2020}, defined as
\begin{equation}
\CHfull = \min_{V\in S^{d},X\in \chi
_{\theta }}[\mathrm{tr}(WV) \vert V \geq Z(X)], 
\label{eq:HCRBdef}
\end{equation}
where $Z(X)$ is a $d\times d$ Hermitian matrix and its matrix elements are defined as\footnote{Notice that in Eq.~\eqref{eq:ZmatDEF} we are actually considering the transposed matrix 
of the one that is usually given in the definition
of the HCRB in the literature, namely $Z(X)_{jk}=\text{Tr}[\hat{\rho}_{\theta }\hat{X}_{j}\hat{X}_{k}]$; this
is in fact equivalent, as the inequality constraint in \eqref{eq:HCRBdef} is invariant under transposition.}
\begin{align}
Z(X)_{jk}=\text{Tr}[\hat{X}_{j}\hat{\rho}_{\theta }\hat{X}_{k}]\,,
\label{eq:ZmatDEF}
\end{align}
in terms of a set of $d$ Hermitian operators $\hat{X}_{j}$ that 
satisfy the local unbiasedness conditions
\begin{eqnarray}
\Tr \left[ \hat{\rho}_{\theta }\hat{X}_{j}\right] &=& 0  \label{eq:Xtracezero} \\
\Tr \left[ \hat{X}_{j} \frac{\partial \hat{\rho}_{\theta }}{\partial \theta_{k}} \right] &=&\delta _{jk}.  \label{eq:Xluc}
\end{eqnarray}
Typically, the HCRB is tighter than both the SLD-CRB and RLD-CRB; indeed, the bounds introduced so far satisfy the following chain of inequalities
\begin{eqnarray}
\tr \left[ W \Sigma _{\theta }(\check{\theta}) \right]&\geq &C^{\mathrm{MI}}(\hat{%
\rho}_{\theta },W)\geq \CHfull  \notag \\
&\geq &\max [\CSfull,C_{\theta }^{\mR}(\hat{%
\rho}_{\theta },W)].  \label{14}
\end{eqnarray}

The discrepancy between the HCRB and the SLD-CRB is limited to a factor two, since we have that~\cite{Tsang2019,Carollo2019}
\begin{eqnarray}
\CHfull \leq \CUfull &\leq &(1+R)\CSfull\leq 2 \CSfull,  \label{15}
\end{eqnarray}
where the first upper bound is defined as
\begin{align}
\CUfull = \CSfull +\left \Vert \sqrt{W}(J^{S})^{-1} \mathcal{I} (J^{S})^{-1}\sqrt{W}\right \Vert_1
\label{eq:HCRBupperbound}
\end{align}
in terms of the mean Uhlmann curvature or asymptotic incompatibility matrix $\mathcal{I}$, with elements
\begin{align}
\mathcal{I}_{jk}= -\frac{i}{2} 
\Tr \left[\hat{\rho}_{\theta }[\hat{L}_{j}^{\mS},\hat{L}_{k}^{\mS}] \right]\,,
\label{eq:UhlmannDef}
\end{align}
and the intermediate bound involves the asymptotic incompatiblity parameter $R=\left \Vert i(J^{S})^{-1}\mathcal{I}\right \Vert
_{\infty }$~\cite{Carollo2019,Razavian2020,Candeloro2021c} (the operator norm $\left \Vert A\right \Vert _{\infty }$ corresponds to the largest singular value of the matrix $A$).

It is worth emphasizing that the SLD-CRB is completely tight, i.e. $C^{\mathrm{MI}}_{\theta}(W) = \CSfull$, when $[\hat{L}_{j}^{S},\hat{L}_{k}^{S}]=0, \, \forall j,k$ and the common eigenstates of these commuting SLD operators are the optimal measurement basis.
Moreover, the SLD-CRB is asymptotically tight, i.e. the less stringent condition $\CHfull = \CSfull$, when the SLDs commute on average $\mathcal{I}_{jk}\equiv 0,$ $\forall j,k$.

\subsection{ Gaussian quantum systems}
\label{s:GaussianCV}

In this section, we briefly summarize the formalism to describe bosonic Gaussian quantum systems~\cite{Serafini2023,Genoni2016,Braunstein2005,Weedbrook2012}.
Let $\hat{r}$=$\left( \hat{x}_{1},\hat{p}_{1},...,\hat{x}_{m},\hat{p}_{m}\right) ^{\! \top}$ be a vector of quadrature operators, describing a $m$-mode continuous variable (CV) quantum system.
The corresponding canonical commutation relations (CCR) can be compactly given by
\begin{equation}
\left[  \hat{r},\hat{r}^{\! \top}\right]=i\Omega ,
\label{eq:sympl_mat}
\end{equation}%
where we have introduced the symplectic matrix
\begin{equation}
    \quad \Omega = \oplus _{j=1}^{m}\Omega _{1}\,, \quad \Omega _{1}=
\begin{pmatrix}
0 & 1 \\
-1 & 0
\end{pmatrix}.
\end{equation}
For the sake of conciseness, we employ natural units ($\hbar = 1$).
The quantum state $\hat{\rho}$ of a CV system can equivalently be described in phase space, in terms of its characteristic function $\chi_{\hat{\rho}}(r)$, which for an operator $\hat{O}$ is defined as~\cite{Serafini2023}
\begin{equation}
    \chi_{\hat{O}}(r) = \Tr \left[ \hat{O} \hat{D}_{-r} \right] 
\end{equation}
and $\hat{D}_{-r}=e^{-i r^{\! \top} \Omega \hat{r} }$ is the displacement operator, while $r$=$\left(
x_{1},p_{1},...,x_{m},p_{m}\right) ^{\! \top}$ is a vector of $2m$ real coordinates in phase-space.
In particular, for a Gaussian state $\hat{\rho}_G$ the characteristic function is Gaussian:
\begin{equation}
\chi_{\hat{\rho}_G}(r)=\exp \left[ -\frac{1}{4}\tilde{r}^{\! \top}\sigma \tilde{r}+i\tilde{r}%
^{\! \top}d\right] , 
\end{equation}%
where $\tilde{r}$=$\Omega r$, $d=\Tr\left[ \hat{r} \hat{\rho} \right]$ is the first moment vector, and $\sigma$ is the covariance matrix
\begin{equation}
\sigma = \Tr\left[ \hat{\rho}\left \{ (\hat{r}-d),(\hat{r}-d)^{\! \top}\right
\} \right] \; .
\label{eq:CMdef}
\end{equation}
Thus, Gaussian states are completely characterized by the first moment $d$ and the covariance matrix $\sigma$.

Gaussian unitaries, generated by Hamiltonians at most quadratic in $\hat{r}$, correspond to symplectic transformations $S \in \mathrm{Sp}_{(2m,\mathbb{R})}$, which are those preserving the symplectic matrix $S^{\! \top} \Omega S = \Omega$.
The action on the first moment vector and on the covariance matrix are $d\mapsto Sd$ and $\sigma \mapsto S\sigma S^{\! \top}$.
Any Gaussian state can be decomposed into normal modes, which are those obtained by a symplectic diagonalization of the covariance matrix, i.e. $\sigma = S^{\! \top} \nu S$ with a diagonal $\nu =\mathrm{diag}(\nu_1,\nu_1, \dots \nu_{m},\nu_{m})$ that contains the $m$ symplectic eigenvalues $\nu_k \geq 1$.
A symplectic eigenvalue equal to one corresponds to a pure normal mode, the number of symplectic eigenvalue strictly greater than unity is known as symplectic rank~\cite{Adesso2006b}.

\bigskip
\textbf{Vectorization procedure.}
We introduce here the notation $\text{vec}\left[ A \right]$ to describe the vectorization of a matrix $A$, i.e. the column vector constructed from columns of the matrix:
\begin{eqnarray}
A &=&
\begin{pmatrix}
a & b \\
c & d%
\end{pmatrix}
\quad \mapsto \quad
\text{vec}\left[ A\right] =
\begin{pmatrix}
    a \\
    c \\
    b \\
    d
\end{pmatrix} \, .
\end{eqnarray}
The following property that relates vectorization with Kronecker products will be used in the derivations
\begin{align}
    \text{vec}\left[ABC\right] = \left( C^{\! \top} \otimes A \right) \text{vec}\left[B\right] \,.
    \label{eq:vectorproduct}
\end{align}
While this vectorization procedure is often applied directly to Hilbert-space operators for finite-dimensional systems, we will apply it to matrices related to the phase-space description of CV systems, i.e. the covariance matrix and the matrix defining the quadratic component of operators.

\section{Results}
\subsection{Overview of main results}
\label{sec:overview}
We here report the main results of our work, leaving further details on the derivations and some examples in the next sections and appendices. 

Given a Gaussian state with covariance matrix $\sigma_{\theta}$, the following block-diagonal matrix
\begin{eqnarray}
    S_{\theta } &=&\frac{1}{2}
    \begin{pmatrix}
    \sigma_\theta  -i\Omega & 0 \\
    0 & \left( \sigma_\theta -i\Omega \right) \otimes \left( \sigma_\theta  -i\Omega \right)%
    \end{pmatrix} ,  \label{eq:Stheta} 
\end{eqnarray}
represents the RLD inner product between zero-mean observables at most quadratic in $\hat{r}$, while its real part
\begin{eqnarray}
    \Re ( S_{\theta } )  &=&\frac{1}{2}
    \begin{pmatrix}
    \sigma_\theta & 0 \\
    0 & \sigma_\theta \otimes \sigma_\theta  - \Omega \otimes \Omega 
    \end{pmatrix} ,  \label{eq:ReStheta} 
\end{eqnarray}
represents the SLD inner product.
These two matrices are the central objects of this work, used to express and compute the different bounds.
In the following, we further assume a Gaussian state with full symplectic rank $m$, i.e. without pure normal modes, so that these matrices can be inverted; at the end of Sec.~\ref{sec:derivation} we will discuss how to relax this assumption.
Introducing vectors containing the derivatives of the first moments and covariance matrix as
\begin{equation}
    \label{eq:Dj_vec}
\bar{D}_j = \begin{pmatrix}
    \frac{\partial d_\theta}{\partial \theta_j}  \\
    \frac{1}{2} \mathrm{vec}\left[ \frac{\partial \sigma_\theta}{\partial \theta_j} \right]
    \end{pmatrix},
\end{equation}
the SLD-QFIM elements can be expressed as
\begin{eqnarray}
    \label{eq:JS_DjDk}
    J_{jk}^{\mS}  = \Re(\Tr[\hat{L}_{j}^{\mS}\hat{\varrho}_{\theta }%
    \hat{L}_{k}^{\mS}]) = \bar{D}_j^{\! \top} \, \mathrm{Re}\left(S_{\theta}\right)^{-1} \bar{D}_k ,  \label{eq:SLD-QFIelements}
\end{eqnarray}
or more explicitly
\begin{align}
    J_{jk}^{\mS} &=\frac{1}{2}\text{vec}\left[ \left. \partial \sigma_\theta \right/ \partial
    \theta _{j}\right] ^{\! \top}(\sigma_\theta \otimes \sigma_\theta -\Omega \otimes \Omega )^{-1}%
    \text{vec}\left[ \left. \partial \sigma_\theta \right/ \partial \theta _{k}\right] \notag \\
    &\,\,\,\,\,\,
    +2(\left. \partial d_\theta \right/ \partial \theta _{j})^{\! \top}\sigma_{\theta} ^{-1}(\left.
    \partial d_{\theta} \right/ \partial \theta _{k}) \,.
\end{align}
Similarly, the RLD-QFIM elements can be expressed as
\begin{eqnarray}
    \label{eq:JR_DjDk}
    J_{jk}^{\mR} &=&\Tr(\hat{L}_{j}^{\mR\dagger }\hat{\rho}_{\theta }%
    \hat{L}_{k}^{\mR}) = \bar{D}_j^{\! \top} \, S_{\theta}^{-1} \bar{D}_k \, ,
\end{eqnarray}
explicitly written as
\begin{align}
J_{jk}^{\mR} &=  \frac{1}{2}\text{vec}\left[ \frac{\partial \sigma_\theta}{\partial
\theta _{j}}\right] ^{\! \top} \left[ \left( \sigma_\theta - i\Omega \right) \otimes \left(
\sigma_\theta -  i\Omega \right) \right] ^{-1}\text{vec}\left[ \frac{\partial \sigma_\theta}{\partial \theta _{k}}\right] \notag \\
&\,\,\,\,\, + 2(\left. \partial d_\theta \right/ \partial
\theta _{j})^{\! \top}(\sigma_\theta -i\Omega )^{-1}(\left. \partial d_\theta \right/ \partial
\theta _{k}). 
\end{align}
By collecting vectors as columns of a matrix 
\begin{equation}
    \bar{D} = \begin{pmatrix} 
        \bar{D}_1 & \dots & \bar{D}_p
    \end{pmatrix},
\end{equation}
we can also compactly express the whole SLD-QFIM and RLD-QFIM as
\begin{equation}
    \label{eq:JS_Dmat}
    J^{\mS} = \bar{D}^{\! \top} \, \mathrm{Re}\left(S_{\theta}\right)^{-1} \bar{D}, \quad J^{\mR} = \bar{D}^{\! \top} \, S_{\theta}^{-1} \bar{D} \, .
\end{equation}

While analogous formulas for the SLD and RLD QFI matrices have already been discussed in the literature~\cite{Monras2013,Gao2014a,Serafini2023,Nichols2018,Safranek2019}, this approach has the merit of showing how the bounds can be understood in a unified framework, and it presents their relation with first moments and covariance matrix quite transparently.

More importantly, our main result is that through the matrix $S_\theta$ it is also possible to efficiently evaluate the HCRB, as follows
\begin{equation}
\label{eq:HCRBSDPform}
\begin{aligned}
\CHfull &=  \underset{V \in \mathbb{S}^{p},\bar{X} \in \mathbb{R}^{{z}{\times}p} }{\text{minimize}} & & \tr \left[W  V  \right]\\
& \quad \text{subject to} & & \begin{pmatrix} V & \bar{X}^{\! \top} R_{{\theta}}^\dag \\
R_{{\theta}} \bar{X} & \mathbb{I}_r \end{pmatrix} \geq 0 \\ 
& & & \bar{X}^{\! \top} \bar{D} = \mathbb{I}_{p}.\,
\end{aligned}\;,
\end{equation}
with $z=2m(1+2m)$, and where $R_\theta$ is a $r{\times}z$ decomposition of the RLD inner product: $S_\theta = R_\theta^\dag R_\theta$, e.g. the square root $R_\theta=\sqrt{S_\theta}$ (with $r=z$).
The structure of this optimization problem is completely analogous to the finite-dimensional case~\cite{Albarelli2019}, and the optimization~\eqref{eq:HCRBSDPform} can be readily recognized as an SDP~\cite{Boyd,Skrzypczyk2023}, a class of convex minimization problem that can be efficiently solved numerically using readily available solvers with a guarantee of global optimality.

\subsection{Derivations}
\label{sec:derivation}
In this section we will give some details on the derivation of the main results we have just outlined, leaving the most technical details to the appendices, and we will also present some additional minor results that we believe will turn useful when discussing quantum estimation problems involving Gaussian quantum states.

% \subsubsection{RLD inner product of quadratic operators}

\bigskip
\textbf{RLD inner product of quadratic operators.}
As before, we will consider a Gaussian quantum state $\hat{\varrho}_\theta$ described by a first moment vector $d_{\theta}$ and a covariance matrix $\sigma_\theta$, with a smooth parametric dependence on the vector of real parameters $\theta$.
We will deal with operators that are at most linear and quadratic in the operators $\hat{r}$, so that we can write them as
\begin{align}
\hat{A} &= {a}^{(0)}\hat{\mathbb{I}} + ({a}^{(1)})^{\! \top} \hat{r} + \hat{r}^{\! \top}{a}^{(2)}\hat{r}\,, \label{eq:quadraticoperator}
\end{align}
where ${a}^{(0)}$ is a scalar complex number, ${a}^{(1)} \in \mathbb{C}^{2m}$, and ${a}^{(2)}$ is a $2m \times 2m$ complex symmetric matrix. 
Notice that we can restrict to symmetric matrices without loss of generality, since a matrix can be decomposed into symmetric and antysymmetric parts and the latter gives rise to terms proportional to $[\hat{r}_j, \hat{r}_k]$, which, because of the CCR, are either zero or proportional to the identity and thus can be included in the complex scalar ${a}^{(0)}$.
Given two generic operators $\hat{A}$ and $\hat{B}$ of this form the expectation value of their product $\Tr [\hat{B}^\dag \hat{\varrho} \hat{A}]$ over a Gaussian state $\hat{\varrho}$ with first moment $d$ and covariance matrix $\sigma$ (known as the RLD inner product~\cite{Hayashi2017c} between the two operators) can be written as a function 
\begin{align}
\Tr[\hat{B}^\dag \varrho_\theta \hat{A}] &= \mathcal{F}(d,\sigma,{a}^{(0)},{a}^{(1)},{a}^{(2)},{b}^{(0)},{b}^{(1)},{b}^{(2)})\,.
\end{align}
We have derived its general expression in Appendix~\ref{a:traceproductformula} and reported it in Eq.~\eqref{eq:traceproductoperators}.
While we do not report it here in the main text, we believe that this relatively compact formula is going to be useful beyond the scope of this work, and we consider it as one of our main results.

In what follows we will actually restrict to operators that are at most quadratic in the canonical operators, but that are also \emph{zero-mean}, that is they satisfy $\Tr[\hat{\varrho}_\theta \hat{A}]=0$.
It is convenient to express operators in the \emph{central basis} defined by the quantum state $\hat{\varrho}_\theta$, as follows 
\begin{align}
\hat{A} =&-\frac{1}{2}\tr[A^{(2)}\sigma_\theta ] \,\hat{\mathbb{I}}%
+ A^{\left( 1\right) \! \top}\left( \hat{r}-d_\theta \hat{\mathbb{I}} \right)  \notag
\\
&+\left( \hat{r}-d_\theta \hat{\mathbb{I}} \right) ^{\! \top}A^{\left( 2\right) }\left( \hat{r}%
-d_\theta  \hat{\mathbb{I}} \right) ,  \label{eq:centralbasis}
\end{align}
where it is easy to show that $A^{(2)} = a^{(2)}$ and $A^{(1)} = a^{(1)} + \left( a^{(2) \! \top} + a^{(2)} \right) d_{\theta}$, while the scalar term is fixed by the zero-mean constraint.
One can write any Gaussian state $\hat{\varrho}_\theta$ with first moment vector $d_{\theta}$ as $\hat{\varrho}_\theta = \hat{D}_{d_{\theta}}^\dag \varrho_\theta^0 \hat{D}_{d_{\theta}}$, where $\hat{D}_d = \exp(i d^{\! \top} \Omega \hat{r})$ denotes the Weyl (displacement) operator, and where $\varrho_\theta^0$ is a Gaussian state with zero first moments and covariance matrix $\sigma_{\theta}$.
Given two zero-mean operators $\hat{A}$ and $\hat{B}$, by exploiting the displacement operator property $\hat{D}_{d_{\theta}} \hat{r} \hat{D}_{d_\theta}^\dag = \hat{r} + d_{\theta} \hat{\mathbb{I}}$, one can  observe that
\begin{align}
\Tr[\hat{B}^\dag \varrho_\theta \hat{A}] = \Tr[\hat{Y}^\dag \varrho_\theta^0 \hat{X}] \,,
\label{eq:TrAB_corrisp_TrXY}
\end{align}
where the operators $\hat{X}=\hat{D}_{d_{\theta}} \hat{A} \hat{D}_{d_\theta}^\dag$ and $\hat{Y}=\hat{D}_d \hat{B} \hat{D}_d^\dag$ can be readily written as in Eq.~\eqref{eq:quadraticoperator}, with corresponding vector and matrices: 
\begin{align}
x^{(0)} &= -\frac{1}{2} \text{tr}[A^{(2)}\sigma]\,, \quad {x}^{(1)} = A^{(1)} \,, \quad {x}^{(2)} = A^{(2)} \,, \\
y^{(0)} &= -\frac{1}{2} \text{tr}[B^{(2)}\sigma]\,, \quad {y}^{(1)} = B^{(1)} \,, \quad {y}^{(2)} = B^{(2)}\,.
\end{align}
Since the quantum state $\varrho_\theta^0$ has zero first moments, we can thus exploit the general formula reported in Eq.~\eqref{eq:traceproductoperators} by setting $d=0$, and by replacing the correct matrices and vectors identified above, and by exploiting the properties and relationships described at the end of Sec.~\ref{s:GaussianCV}, one shows that the RLD inner product between zero-mean quadratic operators can be elegantly rewritten as 
\begin{align} 
\Tr \left[ \hat{B}^\dag \varrho_\theta \hat{A}  \right] = 
\bar{B}^\dag \,
S_\theta
\bar{A}
\label{eq:RLDinnerproduct}
\end{align}
where the matrix $S_\theta$ was introduced in Eq.~(\ref{eq:Stheta}), and where
\begin{equation}
\bar{A} = \begin{pmatrix} A^{(1)} \\ \mathrm{vec} [ A^{(2)} ] \end{pmatrix} \qquad
\bar{B} = \begin{pmatrix} B^{(1)} \\ \mathrm{vec}[ B^{(2)} ] \end{pmatrix} \label{eq:barvector}
\end{equation}
The formula in Eq.~\eqref{eq:RLDinnerproduct} is the basis of most of the derivation we will describe below, and we will refer to it as the RLD inner product equation.

% \subsubsection{SLD operators and SLD-CRB}

\bigskip
\textbf{SLD operators and SLD-CRB}.
The SLD operators of Gaussian states are at most quadratic in the canonical operators, and this fact has been exploited in the literature to obtain explicit formulas for them and for the SLD-QFIM~\cite{Monras2013,Gao2014a,Serafini2023,Nichols2018,Safranek2019}.
For the sake of completeness we report here these results.
Starting with the ansatz that the SLD operator is at most quadratic in the canonical operators, we can write it in the standard basis in the form of Eq.~\eqref{eq:quadraticoperator}
\begin{align}
\hat{L}_{j}^{\mS}=l_{j}^{\mS(0)}\hat{\mathbb{I}}+\left( l_{j}^{\mS(1)}\right) ^{\! \top}%
\hat{r}+\hat{r}^{\! \top}l_{j}^{\mS(2)}\hat{r} \,.
\label{eq:SLDquadratic}
\end{align}
This ansatz is enough to solve the defining equation of the SLD~\eqref{eq:SLDoperator}, and the solution is
\begin{align}
l_j^{\mS(0)} &= -\frac{1}{2} \tr\left[\sigma l_{j}^{\mS(2)}\right]-
l_{j}^{\mS(1) \! \top}d   -d^{\! \top} l_{j}^{\mS(2)} d \label{eq:SLDstandardbasis0} \\
l_j^{\mS(1)} &= 2\sigma ^{-1}\frac{\partial d }{ \partial \theta_{j} }-2\,l_{j}^{\mS(2)}d \label{eq:SLDstandardbasis1} \\
\text{vec}\left[l_j^{\mS(2)}\right] &= 
(\sigma \otimes \sigma
-\Omega \otimes \Omega )^{-1}\,\text{vec}[\left. \partial \sigma \right/
\partial \theta _{j}]
\label{eq:SLDstandardbasis2}
\end{align}
This can be obtained by exploiting the correspondence between phase space differential operators acting on the characteristic function and operators acting on the infinite dimensional Hilbert space~\cite{Serafini2023}. 
These techniques and identities are described in Appendix~\ref{app:correspondence}.
By observing that the SLD operators are zero-mean, we can then write them in the central basis as in Eq.~\eqref{eq:centralbasis}, and obtain the corresponding vectors and matrices
\begin{align}
L_j^{\mS(1)} &= 2\sigma ^{-1}  \frac{\partial d}{\partial \theta
_{j}} \label{eq:SLDcentralbasis1}\\
L_j^{\mS(2)} &= \left( \sigma \otimes \sigma
-\Omega \otimes \Omega \right)^{-1}\,\text{vec} \left[ \frac{\partial \sigma }{
\partial \theta_{j}} \right] \, , \label{eq:SLDcentralbasis2}
\end{align}
which can be compactly written as
\begin{equation}
\label{eq:SLD_Sinv}
\bar{L}^{\mS}_j = \mathrm{Re}[S_{\theta}]^{-1} \begin{pmatrix}
\frac{\partial d}{\partial \theta_j}  \\
\frac{1}{2} \mathrm{vec}\left[ \frac{\partial \sigma}{\partial \theta_j} \right]
\end{pmatrix} = \mathrm{Re}[S_{\theta}]^{-1}  \bar{D}_j
\end{equation}
in terms of the vector $\bar{D}_j$ introduced in Eq.~\eqref{eq:Dj_vec}.
Since the SLD operators are Hermitian these vectors are real-valued and, applying the RLD inner product formula \eqref{eq:RLDinnerproduct}, we can readily write the corresponding SLD-QFI elements as in Eq.~\eqref{eq:SLD-QFIelements}, that we report again here
\begin{align}
    J_{jk}^{\mS} =\Re(\text{Tr}[\hat{L}_{j}^{\mS}\hat{\varrho}_{\theta }%
\hat{L}_{k}^{\mS}])  =\left( \bar{L}_{j}^{\mS}\right) ^{\! \top}\Re\left( S_{\theta
}\right) \bar{L}_{k}^{\mS}\,,
\end{align}
where we have defined the vectors $\bar{L}_j^\mS$ as in Eq.~\eqref{eq:barvector}.
By inserting Eq.~\eqref{eq:SLD_Sinv} we obtain the expressions previously reported in Eqs.~\eqref{eq:JS_DjDk} and~\eqref{eq:JS_Dmat}.

Similarly, by noticing that 
\begin{align}
\text{Tr}[\varrho_\theta [\hat{L}_j^\mS,\hat{L}_k^S]]=-2i \,\Im\left(\text{Tr}[\hat{L}_j^\mS\varrho_\theta\hat{L}_k^\mS]\right) \,,
\end{align}
we can evaluate readily the matrix elements of the mean Uhlman matrix in Eq.~\eqref{eq:UhlmannDef}
\begin{align}
\mathcal{I}_{jk} &= -\frac{i}{2} \text{Tr}[\varrho_\theta [\hat{L}_j^\mS,\hat{L}_k^\mS]] = - \left( \bar{L}_{j}^{S}\right) ^{\! \top}\Im\left( S_{\theta}\right) \bar{L}_{k}^{S}\,,
\label{eq:UhlmannMatrixElements}
\end{align}
that, written compactly as a function of the derivatives of first and second moments of the Gaussian state, reads
\begin{equation}
    \label{eq:UhlmannMatrixFull}
    \mathcal{I} = - \bar{D}^{\! \top} \Re(S_{\theta})^{-1} \Im(S_\theta) \Re(S_{\theta})^{-1} \bar{D} . 
\end{equation}

The actual commutator between the SLD operators can also be derived by exploiting the fact that they are quadratic in the canonical operators as in Eq.~\eqref{eq:SLDquadratic} and the CCR, and expressed in terms of $d_{\theta}$, $\sigma_{\theta}$ and the matrices $L_j^{\mS(2)}=l_j^{\mS(2)}$:
\begin{align}
& [\hat{L}_j^{\mS}, \hat{L}_k^{\mS}] = 4i \left( \partial _{j}d\right) ^{\! \top}\sigma ^{-1}\Omega \sigma
^{-1}\partial _{k}d_\theta  \, \hat{\mathbb{I}} \label{eq:SLDcommutator}  \\
&\, + 4i\left( \left( \partial _{j}d_\theta \right) ^{\! \top}\sigma_\theta
^{-1}\Omega L_{k}^{\mS (2)}-\left( \partial _{k} d_\theta \right) ^{\! \top}\sigma_\theta ^{-1}\Omega
L_{j}^{\mS (2)}\right) \left( \hat{r}-d_\theta\right) \notag \\
&\, + 2 i (\hat{r}-d_\theta)^{\! \top}\left( L_{j}^{\mS (2)}\Omega L_{k}^{\mS (2)}-L_{k}^{\mS (2)}\Omega
L_{j}^{\mS (2)}\right) \left( \hat{r}-d_\theta \right)   \,. \nonumber
\end{align}
We recall that when the $p(p-1)/2$ independent elements in Eq.~\eqref{eq:UhlmannMatrixElements} are zero, then the parameters are asymptotically compatible, while the stronger condition that the actual commutator in Eq.~\eqref{eq:SLDcommutator} is zero ensures that they can be jointly estimated with a precision equal to the SLD-CRB even at the single-copy level.
One can immediately see that the two conditions are equivalent for parameters encoded only in the first moments, while the full commutativity of SLDs is a stronger condition to satisfy when the parameter dependence is also in the covariance matrix.

% \subsubsection{RLD operators and RLD-CRB}
%\label{a:RLDderivation}

\bigskip
\textbf{RLD operators and RLD-CRB.}
A similar approach can be employed also to obtain RLD operators.
This has already been pursued in Ref.~\cite{Gao2014a} using a different notation; we report the results here for completeness and to see the formal analogy with the SLD case.
As previously done for the SLD operators, we can also assume that the RLD operators corresponding to a Gaussian state $\hat{\varrho}_\theta$ are quadratic in the canonical operators, and thus we can write them as
\begin{align}
    \hat{L}_j^{\mR} = l_{j}^{\mR(0)}\hat{\mathbb{I}}+\left( l_{j}^{\mR(1)}\right) ^{\! \top}%
\hat{r}+\hat{r}^{\! \top}l_{j}^{\mR(2)}\hat{r}
\label{eq:RLDquadratic}
\end{align}
or alternatively, in terms of the central basis, as
\begin{align}
    \hat{L}_j^{\mR} &= -\frac{1}{2}\tr[L_j^{\mR (2)}\sigma \text{]}\,\hat{\mathbb{I}}%
+\left( L_j^{\mR \left( 1\right) }\right) ^{\! \top}\left( \hat{r}-d\right)  \notag
\\
&+\left( \hat{r}-d\right) ^{\! \top}L_j^{\mR \left( 2\right) }\left( \hat{r}%
-d\right) ,  \label{eq:RLDcentralbasis}
\end{align}
By exploiting the definition of RLD operators in Eq.~\eqref{eq:RLDoperator}, we can derive the corresponding vectors and matrices for the standard basis
\begin{align}
l_j^{\mR(0)} &= -\frac{1}{2}\tr \left[ (\sigma_\theta -i\Omega
)l_{j}^{\mR(2)}\right] \notag \\
&\,\,\,\,\,\,\, -l_{j}^{\mR(1)^{\! \top}}d_\theta-d_\theta^{\! \top} l_{j}^{\mR(2)} d_\theta \,, \label{eq:RLDstandardbasis0} \\
l_j^{\mR(1)} &= 2(\sigma_\theta -i\Omega )^{-1}  \frac{\partial d_\theta }{ 
\partial \theta_{j} } \nonumber \\
&\,\,\,\,\,\,\,\
 - \left( l_{j}^{\mR(2)} + l_{j}^{\mR(2) \top} \right) d_\theta  \label{eq:RLDstandardbasis1} \\
\text{vec}\left[ l_{j}^{\mR(2)}\right] &= \left[ \left( \sigma_\theta -i\Omega
\right) \otimes \left( \sigma_\theta -i\Omega \right) \right] ^{-1}\text{vec}\left[ \frac{\partial \sigma_\theta}{ \partial \theta_{j}} \right] \label{eq:RLDstandardbasis2}
\end{align}
and for the central basis
\begin{align}
L_{j}^{\mR(1)} &= 2(\sigma_\theta -i\Omega )^{-1}\left. \partial d_\theta \right/
\partial \theta _{j}  \label{eq:RLDcentralbasis1} \\
\text{vec}\left[ L_{j}^{\mR(2)}\right] &= \left[ \left( \sigma -i\Omega
\right) \otimes \left( \sigma_\theta -i\Omega \right) \right] ^{-1}\text{vec}\left[ \frac{\partial \sigma_\theta}{ \partial \theta_{j}} \right]  \,. \nonumber
% \label{eq:RLDcentralbasis2}
\end{align}
These can be compactly written as
\begin{equation}
\label{eq:RLD_Sinv}
\bar{L}^{\mR}_j = S_{\theta}^{-1} \begin{pmatrix}
\frac{\partial d}{\partial \theta_j}  \\
\frac{1}{2} \mathrm{vec}\left[ \frac{\partial \sigma}{\partial \theta_j} \right]
\end{pmatrix} = S_{\theta}^{-1}  \bar{D}_j
\end{equation}
in terms of the vector $\bar{D}_j$ introduced in Eq.~\eqref{eq:Dj_vec}.

More details on this derivation can be found in Appendix~\ref{a:RLDderivation}.
Also in this case we can straightforwardly exploit the RLD inner product formula in Eq.~\eqref{eq:RLDinnerproduct} to obtain the RLD-QFI matrix elements 
\begin{equation}
J_{jk}^{\mR}  = \Tr(\hat{L}_{j}^{\mR \dagger }\hat{\rho}_{\theta }%
\hat{L}_{k}^{\mR})   =(\bar{L}_{j}^{\mR})^{\dagger}  {S}_{\theta }\bar{L}_{k}^{\mR}\,,
\label{eq:RLD-QFIelements}
\end{equation}
where we have defined the vectors $\bar{L}_j^{\mR}$ as in Eq.~\eqref{eq:barvector}; one should notice that in this case the RLD operators $\hat{L}_j^{\mR}$ are still zero-mean (as it can be easily proven from their definition), but in general they are not Hermitian.
By inserting Eq.~\eqref{eq:RLD_Sinv} we obtain the expressions previously reported in Eqs.~\eqref{eq:JR_DjDk} and~\eqref{eq:JS_Dmat}.

% \subsubsection{Evaluation of the HCRB as a semidefinite program}

\bigskip
\textbf{Evaluation of the HCRB as a semidefinite program.}
The definition of the HCRB in Eq.~\eqref{eq:HCRBdef} is given in terms of the matrix $Z(X)$ which is the Gram matrix of the Hermitian operators $\{\hat{X}_j\}$ with respect to the RLD inner product; these operators must also satisfy the local unbiasedness conditions~\eqref{eq:Xtracezero} (zero-mean) and~\eqref{eq:Xluc}.
In principle, the optimization should be carried out over operators of any order.
For the moment, we just assume that they can be chosen to be at most quadratic in the canonical operators; we temporarily postpone 
% to Sec.~\ref{subsec:OptimalityQuadratic} 
the proof that this restriction does not affect the value of the HCRB.
These operators can thus be written in the central basis
\begin{align}
\hat{X}_j =&-\frac{1}{2}\tr[ X_j^{(2)}\sigma ] \,\hat{\mathbb{I}}%
+\left( X_j^{\left( 1\right) }\right) ^{\! \top}\left( \hat{r}-d\right)  \notag
\\
&+\left( \hat{r}-d\right) ^{\! \top}X_j^{\left( 2\right) }\left( \hat{r}%
-d\right) . \label{eq:Xcentralbasis} 
\end{align}
As they are Hermitian and zero-mean, we can exploit the RLD inner product to write the $Z(X)$ matrix elements as
\begin{eqnarray}
Z(X)_{jk} &=&\text{Tr}(\hat{X}_{j} \hat{\rho}_{\theta }\hat{X}_{k}) = \bar{X}_{j}^{\! \top} S_{\theta } \bar{X}_{k},  \label{20}
\end{eqnarray}%
where we have defined
\begin{align}
\bar{X}_j &= \begin{pmatrix} X_j^{(1)} \\ \mathrm{vec} \left[ X_j^{(2)} \right] \end{pmatrix} \,.
\end{align}
We can thus rewrite the whole Hermitian matrix $Z(X)$ as%
\begin{equation}
Z(X)=\bar{X}^{\! \top} S_{\theta }\bar{X},  \label{21}
\end{equation}%
where%
\begin{equation}
\bar{X}=
\begin{pmatrix}
X_{1}^{(1)} & ... & X_{p}^{(1)} \\
\text{vec}[X_{1}^{(2)}] & ... & \text{vec}[X_{p}^{(2)}]%
\end{pmatrix},  \label{22}
\end{equation}
so that the matrix inequality in the HCRB defintion~\eqref{eq:HCRBdef} can be represented as $V\geq
\bar{X}^{\! \top}S_{\theta }\bar{X}.$ By resorting to the Schur complement
condition, this matrix inequality can be converted to a linear matrix
inequality%
\begin{eqnarray}
V-\bar{X}^{\! \top}S_{\theta }\bar{X} \geq 0  \Leftrightarrow \begin{pmatrix}
V & \bar{X}^{\! \top} {R}_{\theta }^{\dagger } \\
{R}_{\theta }\bar{X} & \mathbb{I}_r %
\end{pmatrix} \geq 0,  \label{23}
\end{eqnarray}
with $S_{\theta }$=${R}_{\theta }^{\dagger }{R}_{\theta }$, we can choose, e.g., $R_{\theta} = R_{\theta}^\dag = \sqrt{S_{\theta}}$.

The local unbiasedness condition can be written as
\begin{eqnarray}
\delta _{jk} &=&
\Tr \left[ \hat{X}_{j}\partial \hat{\rho}_{\theta
}/\partial \theta _{k}\right]  
= \frac{1}{2} \Tr \left[ \hat{X}_{j} \{ \hat{\rho}_{\theta },\hat{L}%
_{k}^{S}\} \right]  \notag \\
&=&\Re\left[ \Tr[ \hat{X}_{j}\hat{\rho}_{\theta }\hat{L}_{k}^\mS ] \right] = \bar{X}_{j}^{\! \top} \Re\left( {S}_{\theta }\right)
\bar{L}_{k}^{S} \notag  = \bar{X}_{j}^{\! \top} \bar{D}_k \notag    \,, 
\label{eq:localunbias_Gaussian}
\end{eqnarray}%
and thus one can express the constraint of the minimization problem as
\begin{align}
\label{eq:local_unbiased_const_SLDs}
\bar{X}^{\!\top}  \Re( S_{\theta } ) \bar{L}^{\mS}=  \bar{X}^{\!\top}  \bar{D} = \mathbb{I}_{d}.
\end{align}
It is then clear that the evaluation of the HCRB for Gaussian states can be reformulated as we have previously reported in Eq.~\eqref{eq:HCRBSDPform}, which we write again here for completeness:
\begin{equation}
\begin{aligned}
\CHfull &=  \underset{V \in \mathbb{S}^{p},\bar{X} \in \mathbb{R}^{{z}{\times}p} }{\text{minimize}} & & \tr \left[W  V  \right]\\
& \quad \text{subject to} & & \begin{pmatrix} V & \bar{X}^{\! \top} R_{{\theta}}^\dag \\
R_{{\theta}} \bar{X} & \mathbb{I}_r \end{pmatrix} \succeq 0 \\ 
& & & \\bar{X}^{\! \top} \bar{D} = \mathbb{I}_{p}.\,
\end{aligned}\;,
\end{equation}
with $z=2m(1+2m)$, and where $R_\theta$ is a $r{\times}z$ decomposition of the RLD inner product: $S_\theta = R_\theta^\dag R_\theta$, e.g. the hermitian positive semidefinite $\sqrt{S_\theta}$ or the Cholesky decomposition.
The optimization~\eqref{eq:HCRBSDPform} is analogous to the finite-dimensional case~\cite{Albarelli2019} and it corresponds to a SDP~\cite{Boyd,Skrzypczyk2023}, since the objective function is linear and features only linear equality constraints and semidefinite inequality constrains.
SDPs are a class of convex minimization problems that can be efficiently solved numerically with readily available solvers with a guarantee of global optimality.
In practice, one can directly feed this exact formulation to a modeling language for convex optimization, such as CVXPY~\cite{Diamond2016}, which we have used to obtain numerical results.

% \subsubsection{Proof of optimality of quadratic observables}
% \label{subsec:OptimalityQuadratic}

\bigskip
\textbf{Optimality of quadratic observables.}
Let us introduce the (state-dependent) commutation superoperator~\cite{Holevo2011b,Holevo1977}, defined by the operator equation
\begin{equation}
    \label{eq:Dsuper_def}
    \left\{  \Dsuper ( \hat{X}), \hat{\rho}_{\theta} \right\}  = i \left[ \hat{X}  , \hat{\rho}_{\theta} \right].
\end{equation}
This map relates the commutator and the anticommutator with the state and thus also the RLD and SLD inner products.
The commutation superoperator is particularly useful to simplify the computation of the HCRB, since it holds that the search for the optimal $\{X_i\}$ can be restricted to any $\Dsuper$-invariant subspace that also contains $\operatorname{span}_{\mathbb{R}}\{\hat{L}_1^{\mS}, \dots, \hat{L}_p^{\mS} \}$~\cite{Holevo2011b,Demkowicz-Dobrzanski2020}.
The missing step to prove that our formulation of the HCRB is correct is thus to show that the linear space of quadratic observables is invariant under the action of the superoperator $\Dsuper$.
A similar statement was proven for linear observables by Holevo~\cite{Holevo2011b} and indeed can be used to simplify its calculations of the HCRB for parameters encoded in the first moments, see Ref.~\cite{Bradshaw2017} for a recent application.

Let us introduce the notation $\hat{Z}=\Dsuper(\hat{X})$ for simplicity.
We can rewrite Eq.~\eqref{eq:Dsuper_def} in terms of characteristic functions 
\begin{equation}
    \label{eq:Dsuper_chi}
\chi_{( \hat{X} \hat{\rho} - \hat{\rho} \hat{X} )} ( \tilde{\vec{r}} )= i\chi_{( \hat{Z} \hat{\rho} + \hat{\rho} \hat{Z} )} ( \tilde{\vec{r}}).
\end{equation}
While $\hat{X}$ is at most quadratic by assumption, we can now make the ansatz that $\hat{Z}$ is quadratic too and show that this equation indeed always has a solution of this form.

Actually, there is no need to do show this explicitly, since we can just notice that obtaining $\hat{Z}$ is completely analogous to the calculation of the SLD operator for the estimation of a unitarily encoded parameter $\theta$ with a generator Hamiltonian $2 \hat{X}$, i.e. $e^{-i 2 \theta \hat{X} } \hat{\rho}_G e^{ i 2 \theta \hat{X} }$.
Thus, since the SLD operators for parameters encoded in Gaussian states are at most quadratic in the canonical operators~\cite{Monras2013,Serafini2023}, we can immediately conclude that this equation can always be satisfied by an operator $\hat{Z}$ which is also at most quadratic.
This proves the invariance of the subspace of operators that are quadratic polynomials of the canonical variables with respect to the action of the superoperator $\mathcal{D}_{\rho_{G}}$ of a Gaussian state $\hat{\rho}_G$.

For completeness, in App.~\ref{app:explicit_Dsuper} we present the explicit form of Eq.~\eqref{eq:Dsuper_chi} for Gaussian $\hat{\rho}_\theta$ and quadratic $\hat{X}$ and $\hat{Z}$.
From that, it is clear that the solution involves inverting $\Re(S_{\theta})$, just as in the computation of the SLD-QFIM.

% \subsubsection{The SLD-CRB and RLD-CRB as semidefinite programs}

\bigskip
\textbf{SLD-CRB and RLD-CRB as SDPs.}
Similarly to what is shown in \cite{Nagaoka1989} for finite-dimesinonal systems, the SLD-CRB and RLD-CRB can be described in the same minimization form as the HCRB.
However, these minimizations have weaker constraints, from which it immediately follows they must be lower bounds on the HCRB, since the minimization is over a larger set.

The scalar SLD-CRB can be obtained by imposing only
$V \geq \Re[ Z(X) ]$, which is only the real part of the inequality constraint in the HCRB.
Thus, we can evaluate it as follows
\begin{equation}
    \label{eq:CSSDPform}
    \begin{aligned}
    \CSfull &=  \underset{V \in \mathbb{S}^{p},\bar{X} \in \mathbb{R}^{{z}{\times}d} }{\text{minimize}} & & \tr \left[W  V  \right]\\
    & \quad \text{subject to} & & \begin{pmatrix} V & \bar{X}^{\! \top} \tilde{R}_{{\theta}}^{\! \top} \\
    \tilde{R}_{{\theta}} \bar{X} & \mathbb{I}_r \end{pmatrix} \succeq 0 \\ 
    & & & 
    \bar{X}^{\! \top} \bar{D} = \mathbb{I}_{p}.\,
    \end{aligned}\;,
\end{equation}
in terms of the real-valued matrix decomposition $\Re( S_{\theta} ) = \tilde{R}_{{\theta}}^{\! \top}  \tilde{R}_{{\theta}}$.
This is equivalent to the minimization of $\tr \left[ W \Re[ Z(X)] \right]$.

The scalar RLD-CRB can also be obtained in a similar form, by relaxing the constraints that the operators are hermitian:
\begin{equation}
    \label{eq:CRSDPform}
    \begin{aligned}
    \CRfull &=  \underset{V \in \mathbb{S}^{p},\tilde{X} \in \mathbb{C}^{{z}{\times}d} }{\text{minimize}} & & \tr \left[W  V  \right]\\
    & \quad \text{subject to} & & \begin{pmatrix} V & \bar{X}^\mathsf{\dag} R_{{\theta}}^\dag \\
    R_{{\theta}} \bar{X} & \mathbb{I}_r \end{pmatrix} \succeq 0 \\ 
    & & & \bar{X}^{\! \top} \bar{D} = \mathbb{I}_{p}.\,
    \end{aligned}\;,
\end{equation}
where now the optimization is over complex-valued matrices $\tilde{X}$ instead of real ones.

% \subsubsection{Pathological states with pure normal modes}
% \label{subsec:pathological}

\bigskip
\textbf{Pathological states with pure normal modes.}
When one or more of the symplectic eigenvalues are exactly equal to one, and thus the symplectic rank~\cite{Adesso2006b} is smaller than $m$, it means that one or more normal modes of the Gaussian states are pure.
In this case, the matrix $S_\theta$ has some zero eigenvalues and thus it defines a so-called pre-inner product, i.e. there can be operators $\hat{Y}$ for which $\Tr [ \hat{Y}^\dag \hat{\rho}_\theta \hat{Y} ] = 0$ even if $\hat{Y} \neq 0$.
In this case also the matrix $\left( \sigma \otimes \sigma - \Omega \otimes \Omega \right)$ corresponding to the SLD inner product on quadratic observables is not invertible~\cite{Serafini2023}.
In principle, one could define equivalence classes, by considering the quotient space with the kernel of this operator~\cite{Holevo2011b}, similarly to the explicit construction in Ref.~\cite{Albarelli2019} for finite-dimensional rank-deficient states.
We practically solve this problem by noting that this approach is equivalent to regularizing the covariance matrix by adding an arbitrarily small perturbation $\sigma \mapsto (1-\epsilon)\sigma + \epsilon \mathbb{I} $ that makes all the symplectic eigenvalues strictly greater than one~\cite{Safranek2019}.
The analytical limit $\epsilon \to 0$ can be taken at the end of the calculation, or in numerical calculations it is enough to take a small numerical value and checking numerical convergence.

\subsection{Case studies}
\label{sec:examples}

In this section, we will exploit the results we have presented above in order to derive the ultimate bounds on the joint estimation of multiple parameters for two paradigmatic examples: i) the estimation of phase and loss describing the evolution of a single-mode Gaussian state; ii) the estimation of complex displacement and squeezing for single- and two-mode Gaussian states.

In both cases, we evaluate the three bounds---SLD-CRB, RLD-CRB, and HCRB---the latter obtained by solving numerically the semidefinite program introduced above.
Concretely, we have implemented this in Python, using the modeling language CVXPY~\cite{Diamond2016}.

We also compute the commutator between the SLD-operators and the corresponding mean Uhlmann curvature matrix $\mathcal{I}$, in order to discuss the potential achievability of the SLD-CRB.
A notebook containing the code used to generate these figures are available in~\cite{Chang2025a}.

% \subsubsection{Phase and Loss}

\begin{figure}[tbp]
\centering 
\includegraphics{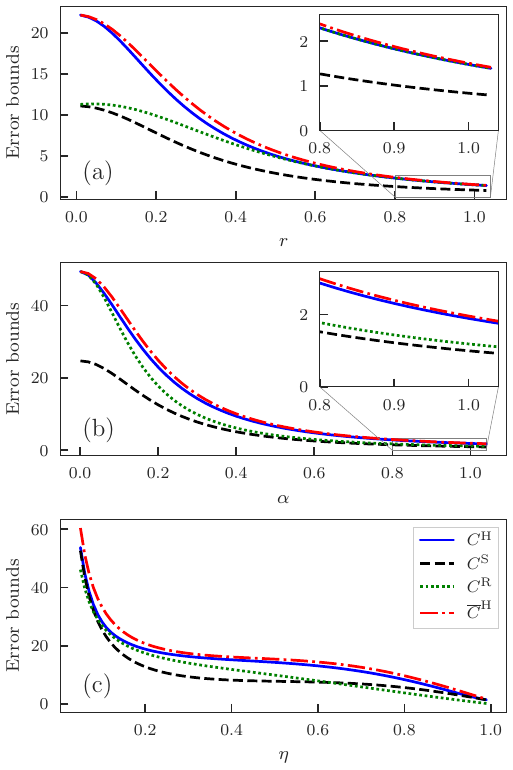}
\caption{(Color online) Error bounds as a function of (a) the squeezing parameter $r$ of the displaced squeezed vacuum state with $\eta =0.5$, $\mathrm{Re}(\alpha )=0.3$, and $\mathrm{Im}(\alpha)=0$, and of (b) the displacement parameter $\alpha $ with $\eta =0.5$, $\mathrm{Im}(\alpha)=0$, and $r=0.2$, and of (c) the photon loss strength $\protect \eta $\ with $r=0.2,$ Re($\protect \alpha $)$=0.3,$
and Im($\protect \alpha $)$=0$.
\label{f:phaseloss}}
\end{figure}

\bigskip
\textbf{Phase and Loss.}
We start by considering the problem of joint estimating a phase $\varphi$ and a loss strength $\eta$, corresponding respectively to a unitary phase rotation in phase space, and to the noisy channel describing the typical photon (or more in general, excitation) loss in bosonic quantum systems.
This is a paradigmatic multiparameter estimation problem, mostly studied in terms of the SLD-CRB~\cite{Pinel2013,Crowley2014,Ragy2016,Nichols2018}, even if the two parameters are not compatible and it requires evaluation of the HCRB~\cite{Albarelli2019,Conlon2020}.
This estimation problem is particularly relevant and interesting in quantum optics, since phase and loss are intrinsically linked in optical samples~\cite{Gianani2021}.
Even when the parameter is just the phase, the inevitable optical losses represent one of the main imperfections of real-life devices, and their amount must be known to implement optimal metrological strategies~\cite{Demkowicz-Dobrzanski2015a}.

Interestingly, the features of the problem depend strongly on the choice of probe state.
For example, when applying phase and loss to a coherent probe state, the output state remains pure and the evolution locally resembles a complex displacement operation in phase space, and thus the problem will share properties with the estimation of orthogonal displacements.
On the other hand, for more general input states, the output state is mixed, making the problem even more interesting and definitely not trivial.

We will consider a single-mode displaced squeezed vacuum state as the input state which is fully characterized by the first moment vector and the covariance
matrix
\begin{align}
d_{\mathrm{in}} &=\sqrt{2}\left(
\begin{array}{c}
\Re(\alpha) \\ \Im(\alpha)%
\end{array}
\right) \,,\\
\sigma _{\mathrm{in}}&= \left(
\begin{array}{cc}
e^{2r} & 0 \\
0 & e^{-2r}%
\end{array}%
\right) \,,
\end{align}
where $\alpha \in \mathbb{C}$ corresponds to the complex amplitude of the 
displacement, while $r \in \mathbb{R}$ is the squeezing parameter.

Then the initial state is sequentially passed through the phase shifter and
the photon loss channel to generate an output state $\varrho_{\mathrm{out}}$ which will depend on the two 
parameters $\varphi$ and $\eta$. In the Gaussian formalism the evolution is
fully determined by the one for first moment vector and covariance
matrix, which reads
\begin{eqnarray}
    d_{\mathrm{out}} &=& \sqrt{\eta} R_{\varphi}d_{\mathrm{in}},  \notag \\
    \sigma _{\mathrm{out}} &=& \eta R_{\varphi }\sigma_{\mathrm{in}} R_{\varphi}^{\! \top} + (1-\eta) \mathbb{I}_2,
    \label{40}
    \end{eqnarray}%
where the matrix describing the phase rotation is
\begin{align}
R_{\varphi}&= 
\begin{pmatrix}
\cos\varphi & \sin\varphi \\
-\sin\varphi & \cos\varphi%
\end{pmatrix} \,.
\end{align}
Notice that the phase parameter can take values $\varphi \in [0,2\pi]$,
while the loss strength is bounded as $\eta\in[0,1]$, where $\eta=1$ corresponds to perfect
transmission (zero loss) and for $\eta=0$ the input state is completely 
replaced by the vacuum state (maximum loss).

In the following we prefer to report analytical results obtained 
for two specific classes of input states, that is coherent states and squeezed vacuum states, and by 
fixing the loss parameter to $\eta=1/2$. We refer to the Appendix~\ref{a:appendix_phaseloss} for other analytical results, obtained for generic values of the loss parameter $\eta$. 

In order to start to characterize the estimation problem, we first 
evaluate the vectors and matrices characterizing the SLD-operators $\hat{L}_\varphi^\mS$
and $\hat{L}_\eta^\mS$ via Eqs.~\eqref{eq:SLDcentralbasis1} and \eqref{eq:SLDcentralbasis2}, 
and the corresponding commutator 
$[\hat{L}_\varphi^\mS ,\hat{L}_\eta^\mS]$ via Eq.~\eqref{eq:SLDcommutator}. 
For a coherent state input (that is, by fixing $r=0$), we find
\begin{align}
\left[ \hat{L}_{\varphi }^{\mS},\hat{L}_{\eta }^{\mS}\right] &= 4i\left \vert \alpha \right \vert ^{2}\,\,\,  (\textrm{for}\,\, r=0\textrm{ and }\eta=1/2)\,,
\end{align}
showing how, in fact, in this case the commutator between SLD operators resembles the one between
the canonical operators $\hat{x}$ and $\hat{p}$. Similarly, for 
an input squeezed vacuum state (that is, by fixing $\alpha=0$) we obtain
\begin{align}
\left[ \hat{L}_{\varphi }^{\mS},\hat{L}_{\eta }^{\mS}\right] &= 
4i\left( \Delta_1 \hat{x}^2 + \Delta_2 \hat{p}^2 + \Delta_3 \{\hat{x},\hat{p}\} \right)
\notag
 \\
&\,\,\,\,\,\,\,\,\,\, (\textrm{for } \alpha=0 \textrm{ and } \eta=1/2), 
\end{align}
where
\begin{eqnarray}
\Delta _{1} &=&\frac{4\left[ \sinh ^{2}2r-4\cos 2\varphi \cosh r\sinh
^{3}r\right] }{(3+\cosh 2r)^{2}},  \notag \\
\Delta _{2} &=&\frac{4\left[ \sinh ^{2}2r+4\cos 2\varphi \cosh r\sinh
^{3}r\right] }{(3+\cosh 2r)^{2}},  \notag \\
\Delta _{3} &=&\frac{16\cosh r\sin 2\varphi \sinh ^{3}r}{(3+\cosh
2r)^{2}}.  \label{42}
\end{eqnarray}%
This result shows how the estimation problem is indeed more complex and interesting, for general input states.
Similarly, we have exploited Eq.~\eqref{eq:UhlmannMatrixElements} to evaluate the matrix elements of the mean Uhlmann curvature.
These can be evaluated analytically, and the off-diagonal elements are in general not zero; as before we report here the results for $\eta=1/2$ and for input coherent state
\begin{align}
\mathcal{I} &=
\begin{pmatrix}
0 & 2\left \vert \alpha \right \vert ^{2} \\
-2\left \vert \alpha \right \vert ^{2} & 0%
\end{pmatrix} \,\, (r=0 \text{ and } \eta=\frac{1}{2}) \, . 
\end{align}
and for input squeezed vacuum state,
\begin{align}
\mathcal{I} &=
\begin{pmatrix}
0 & \frac{8\sinh ^{2}2r}{(3+\cosh 2r)^{2}} \\
-\frac{8\sinh ^{2}2r}{(3+\cosh 2r)^{2}} & 0%
\end{pmatrix} \,\, (\alpha =0 \text{ and } \eta=\frac{1}{2}).
\label{43}
\end{align}
These results clearly show that the SLD-CRB is not achievable, not even in the limit of measurements on an 
asymptotically large number of copies of the output Gaussian states.

We have then exploited the formulas outlined in the previous sections 
to evaluate analytically the SLD-CRB $C_{(\varphi,\eta)}^{\mS}$, the RLD-CRB $C_{(\varphi,\eta)}^{\mR}$, 
and the upper bound to the HCRB $\overline{C}_{(\varphi,\eta)}^{\mathrm{H}}$, (notice that we will consider
the \emph{uniform} weight matrix $W=\mathbb{I}_2$).
As in the previous case, the analytical formulas for the generic output states are quite cumbersome and 
not particularly insightful.
However, by fixing the loss parameter $\eta=1/2$, and for an input coherent state, that is by fixing $r=0$, we obtain
\begin{align}
C_{(\varphi,\eta)}^{\mS} &= 1/\left \vert \alpha \right \vert ^{2} \,, \\
C_{(\varphi,\eta)}^{\mR}&= \overline{C}_{(\varphi,\eta)}^{\mathrm{H}} = 2C_{(\varphi,\eta)}^{\mS} = 2/\left \vert \alpha \right \vert ^{2} \,,
\end{align}
Similarly, still by fixing $\eta=1/2$ and for a squeezed vacuum state, that is for $r=0$, one obtains
\begin{align}
C_{(\varphi,\eta)}^{\mS} &= \csch^{2}r+(\tanh ^{2}r)/4 \,, \\
C_{(\varphi,\eta)}^{\mR}&= \overline{C}_{(\varphi,\eta)}^{\mathrm{H}} = C_{\theta }^{\mS}+\csch^{2}r\,.
\end{align}
For $\eta=1/2$ and for these choices of the initial state parameters, one thus evidently obtains that the 
HCRB is equal to the RLD-CRB and thus
\begin{align}
    C_{(\varphi,\eta)}^{\mathrm{H}} &=
    \biggl\{ 
    \begin{array}{c}
        2 / \left \vert \alpha \right \vert ^{2} \,\,\qquad\,\,\,\qquad\qquad\,\,\,(\textrm{for}\,\, r=0) \\
        2\csch^{2}r+(\tanh ^{2}r)/4 \,\,\,\,  (\textrm{for}\,\, \alpha=0).
    \end{array}
\end{align}
We have to remark that, as we show in the Appendix~\ref{a:appendix_phaseloss}, the result above is confirmed independently on the value of $\eta$ for input coherent states (analogously to the estimation of orthogonal displacements~\cite{Holevo2011b}), while for squeezed vacuum states the HCRB is in general larger than the RLD-CRB.

For generic values of the parameters characterizing the input state, $\alpha$ and $r$, and the quantum channel, $\varphi$ and $\eta$, we had to evaluate the HCRB $C_{(\varphi,\eta)}^{\mathrm{H}}$ numerically by exploiting its SDP formulation.
We present the corresponding  results in Fig.~\ref{f:phaseloss}, comparing the HCRB to the SLD-CRB, the RLD-CRB, and its upper bound, and showing their behaviour as a function of the  different physical parameters.

We observe a clear gap between the HCRB and both the SLD-CRB and the RLD-CRB in general;
% We observe that in general there is a visible gap between the HCRB and both the SLD-CRB and the RLD-CRB; 
on the other hand, we can also observe how the upper bound $\overline{C}_{(\varphi,\eta)}^{\mathrm{H}}$ gives in general a much better approximation of the HCRB, as the difference between the two curves is typically relatively small.
When plotting the results as a function of the loss parameter $\eta$ (see panel (c) of Fig.~\ref{f:phaseloss}) we also find that a pretty good approximation of the HCRB is given respectively by the RLD-CRB for small values of $0<\eta\lesssim 0.25$, and by the SLD-CRB in the large loss regime, i.e. for $\eta\lesssim 1$.

\bigskip
\textbf{Displacement and squeezing estimation.}
We will now consider another paradigmatic case of quantum parameter estimation with Gaussian states, that is the estimation of displacement and squeezing for single- and two-mode states of the form
\begin{align}
\hat{\varrho}^{(1)} &= \hat{D}(\alpha) \hat{S}_1(r) \hat{\nu}_{{n}} \hat{S}_1(r)^\dag \hat{D}(\alpha)^\dag\,, \label{eq:dispequeeze1} \\
\hat{\varrho}^{(2)} &= (\hat{D}(\alpha) \otimes \hat{\mathbb{I}} ) \hat{S}_2(r) (\hat{\nu}_{{n}} \otimes \hat{\nu}_{{n}}) \hat{S}_2(r)^\dag (\hat{D}(\alpha)^\dag \otimes \hat{\mathbb{I}})\,, \label{eq:dispequeeze2}
\end{align}
where we have introduced the displacement operator $\hat{D}(\alpha) = \exp\{\alpha \hat{a}^\dag - \alpha^* \hat{a}\}$, the single- and two-mode squeezing operators $\hat{S}_1(r)=\exp\{(r/2)(\hat{a}^{\dag 2} - \hat{a}^2)\}$, $\hat{S}_2(r) = \exp\{r (\hat{a}^\dag \hat{b}^\dag - \hat{a} \hat{b})\}$, and where $\hat{\nu}_{{n}} = ({n}+1)^{-1}\sum_m ({n}/({n}+1))^m |m\rangle\langle m|$ represents a thermal state with ${n}$ average excitations.
We will consider estimation of the three parameters $\theta = (\Re(\alpha),\Im(\alpha),r)^{\! \top}$, corresponding respectively to real and 
imaginary part of the displacement parameter, and the squeezing parameter.

The estimation of displacement parameters has been studied in great detail~\cite{Yuen1973,Genoni2013b,Bradshaw2017,Bradshaw2017a} and the HCRB has been evaluated in~\cite{Bradshaw2017,Bradshaw2017a}.
By fixing $\bar{n}=0$, the states in Eqs.~(\ref{eq:dispequeeze1}) and \eqref{eq:dispequeeze2} corresponds to pure quantum states, and in particular to displaced single- and two-mode squeezed vacuum states. Remarkably, the HCRB corresponding to the joint estimation of complex displacement and squeezing for such states has been derived analytically in~\cite{Bressanini2024a}.
In the following we will exploit the SDP formulation to evaluate numerically the HCRB also for $\bar{n}>0$, that is in the case of mixed displaced squeezed states.

% \subsubsection{Single-mode displaced squeezed thermal state}

We start by considering single-mode displaced squeezed thermal states in Eq.~(\ref{eq:dispequeeze1}), whose first moment $%
d_{s} $ and covariance matrix $\sigma _{s}$ can be expressed as
\begin{eqnarray}
d_{s} &=&\sqrt{2}
\begin{pmatrix}
\text{Re(}\alpha \text{)} & \Im ( \alpha ) 
\end{pmatrix}^{\! \top},  \notag \\
\sigma _{s} &=&(2n+1)
\begin{pmatrix}
e^{2r} & 0 \\
0 & e^{-2r}%
\end{pmatrix}, 
\label{44}
\end{eqnarray}%
where $n$ is the mean photon number of thermal state. 
As done in the previous example, we will exploit the formulas
described in the previous section to evaluate analytically the commutators
between the different SLD operators, the Uhlmann curvature matrices,
the different bounds, SLD-CRB $C_{\theta }^{\mS}$, RLD-CRB $C_{\theta }^{\mR}$,
and the upper bound $\overline{C}_{\theta }^{\mathrm{H}}$ (as before we will consider the uniform
weight matrix $W = \mathbb{I}_3$); we will then compare these last results  with the HCRB, numerically evaluated via SDP optimization.

The commutation relation between the SLD operators and the Uhlmann curvature matrix 
can be respectively expressed as%
\begin{eqnarray}
\left[ \hat{L}_{\Re\left( \alpha \right) }^{\mS},\hat{L}_{\Im%
(\alpha )}^{\mS}\right] &=&\frac{8i}{(2n+1)^{2}},  \notag \\
\left[ \hat{L}_{\Re\left( \alpha \right) }^{\mS},\hat{L}_{r}^{\mS}\right] &%
=&\frac{8i\Im(\alpha )-4\sqrt{2}i\hat{p}_{1}}{2n(n+1)+1},
\notag \\
\left[ \hat{L}_{\Im\left( \alpha \right) }^{\mS},\hat{L}_{r}^{\mS}\right] &%
=&\frac{8i\Re(\alpha )-4\sqrt{2}i\hat{x}_{1}}{2n(n+1)+1},
\label{45}
\end{eqnarray}%
and%
\begin{equation}
\mathcal{I} =\left(
\begin{array}{ccc}
0 & \frac{4}{(2n+1)^{2}} & 0 \\
-\frac{4}{(2n+1)^{2}} & 0 & 0 \\
0 & 0 & 0%
\end{array}%
\right) .  \label{46}
\end{equation}%

Similarly we can evaluate SLD-CRB, RLD-CRB and the upper bound, that read
\begin{eqnarray}
C_{\theta }^{\mS} &=&\frac{(2n+1)^{3}\cosh (2r) +2n(1+n) +1 }{2(2n+1)^{2}},
\label{eq:dispsqueez1_SLD} \\
C_{\theta }^{\mR} &=&C_{\theta }^{\mS}+\frac{n(n+1)}{2n(n+1)+1},  \label{eq:dispsqueez1_RLD} \\
\overline{C}_{\theta }^{\mathrm{H}} &=&C_{\theta }^{\mS}+\frac{1}{2}.  \label{eq:dispsqueez1_upper}
\end{eqnarray}%
We observe how in this case $C_{\theta }^{\mS}\leq C_{\theta }^{\mR}$, meaning
that the RLD-CRB is always more informative than the SLD-CRB (notice that 
the two bounds coincide for 
$n=0$, such that $C_{\theta }^{\mR}=C_{\theta }^{\mS}=\sinh ^{2}r+1$).
Moreover, the second term of $C_{\theta }^{\mR}$ tends to a constant of $1/2$ when
the value of $n$ is relatively large, so that for by increasing the number of 
thermal photons one obtains $C_{\theta }^{\mR}=C_{\theta }^{\mathrm{H}} = \overline{C}_{\theta}^{\mathrm{H}}$ (notice that eventually for $n\to\infty$, one has $C_{\theta }^{\mS}=C_{\theta }^{\mR}=\overline{C}_{\theta
}^{\mathrm{H}}$, as also the Ulhmann curvature matrix goes to zero: 
in this regime however all these bounds also go to infinity 
as the thermal fluctuations do not allow to estimate efficiently these parameters).

\begin{figure}[tbp]
\includegraphics{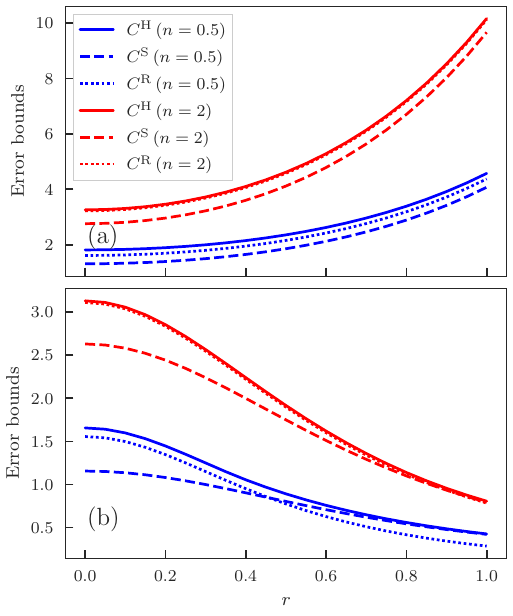}
\caption{(Color online) Error bounds as a function of the squeezing
parameter $r$ for the estimation of complex dispalcement and squeezing by employing 
(a) single-mode displaced squeezed thermal states and (b)
two-mode displaced squeezed thermal states.
We do not show a curve for the upper bound $\overline{C}^{\mathrm{H}}$ because it overlaps with the HCRB in this case.
}
\label{f:dispsqueezing}
\end{figure}

To better understand the behaviour explained above and to compare the 
bounds along with the HCRB, we plot them for two values of the number
of thermal photons, and as a 
function of the squeezing parameter $r$ in Fig.~\ref{f:dispsqueezing}(a) 
(it is worth noting that these three error bounds do not depend on Re($\alpha $) and Im($\alpha $)). 
We start by observing that, as already shown in~\cite{Bressanini2024a} for the 
pure state case, all bounds are monotonically increasing with $r$, meaning that 
squeezing in this case is in general detrimental for the joint estimation of 
all the three parameters.
As regards the relationship between the different bounds, we first find that
in all these example the HCRB $C_{\theta }^{\mathrm{H}}$ numerically coincide with the 
$\overline{C}_{\theta }^{\mathrm{H}}$ reported in Eq.~\eqref{eq:dispsqueez1_upper}.
Moreover, while for small values of thermal photons ($n=0.5$) the HCRB is significantly different from the RLD-CRB, already for $n=2$, the two curves almost coincide, and they differ from the SLD-CRB just for a constant $1/2$.

% \subsubsection{Two-mode displaced squeezed thermal state}
We can now consider the two-mode displaced squeezed thermal state
in Eq.~(\ref{eq:dispequeeze2}). As done in Ref.~\cite{Bressanini2024a}, we actually consider the same state injected into a balanced beam splitter (which can in fact be always considered as a part of the measurement strategy), which generate a tensor product of single-mode squeezed thermal states, corresponding to the following first moment vector and covariance matrix
\begin{eqnarray}
d &=&\left(
\begin{array}{cccc}
\Re\left( \alpha \right)  & \Im(\alpha ) & -\Re\left(
\alpha \right)  & -\Im(\alpha )%
\end{array}%
\right) ^{\! \top},  \notag \\
\sigma &=&(2n+1)\sigma _{\text{I}}\oplus \sigma _{\text{II}},
\label{48}
\end{eqnarray}%
where $\sigma _{\text{I}}$=$
\begin{pmatrix}
e^{2r} & 0 \\
0 & e^{-2r}%
\end{pmatrix} $ and $\sigma _{\text{II}}$=$
\begin{pmatrix}
e^{-2r} & 0 \\
0 & e^{2r}%
\end{pmatrix} .$ 
As done in the previous examples, we can obtain the
corresponding commutation relations and Uhlmann curvature matrix elements, which read
\begin{eqnarray}
\left[ \hat{L}_{\Re\left( \alpha \right) }^{\mS},\hat{L}_{\Im%
(\alpha )}^{\mS}\right]  &=&\frac{8i}{(1+2n)^{2}},  \notag \\
\left[ \hat{L}_{\Re\left( \alpha \right) }^{\mS},\hat{L}_{r}^{\mS}\right]
&=&-\frac{4i\left( \hat{p}_{1}+\hat{p}_{2}\right) }{1+2n(1+n)},  \notag \\
\left[ \hat{L}_{\Im\left( \alpha \right) }^{\mS},\hat{L}_{r}^{\mS}\right]
&=&-\frac{4i(\hat{x}_{1}+\hat{x}_{2})}{1+2n(1+n)},  \label{49}
\end{eqnarray}%
and%
\begin{equation}
\mathcal{I} =
\begin{pmatrix}
0 & \frac{4}{(1+2n)^{2}} & 0 \\
-\frac{4}{(1+2n)^{2}} & 0 & 0 \\
0 & 0 & 0%
\end{pmatrix} . \label{50}
\end{equation}%
Further, one can analytically derive the SLD-CRB, RLD-CRB and the upper bound on the HCRB, i.e.,
\begin{eqnarray}
C_{\theta }^{\mS} &=&\frac{2 {\rm sech}(2r) (1+2n)^{3} + 2n(1+n)+1}{4(2n+1)^{2}},
\label{eq:dispsqueez2_SLD} \\
C_{\theta }^{\mR} &=&\frac{n^{2}(n+1)^{2}}{(2n+1)^{2}\left[ 2n(n+1)+1\right] }
\label{eq:dispsqueez2_RLD} \\
&&+\frac{2n(n+1)}{(2n+1)\cosh (2r)-1},  \notag \\
\overline{C}_{\theta }^{\mathrm{H}} &=&C_{\theta }^{\mS}+\frac{1}{\cosh (4r)+1},  \label{eq:dispsqueez2_upper}
\end{eqnarray}
In this case there is no fixed hierarchy between the RLD-CRB and the
SLD-CRB. For example, for $n$=$0$, one has $C_{\theta }^{\mS}=  \left( 2 \operatorname{sech} (2r)+1 \right) / 4$,
while $C_{\theta }^{\mR}=0$, implying that the RLD-CRB cannot provide useful information for the joint estimation problem.
One can however find values of $n$ and $r$ where the RLD-CRB is more informative, that is $C_{\theta }^{\mR}>C_{\theta }^{\mS}$.
Moreover, we observe that in the limit of large squeezing the SLD-CRB $C_{\theta }^{\mS}$ tends to coincide with the upper bound $\overline{C}_{\theta }^{\mathrm{H}}$, meaning that in this regime it also corresponds to the HCRB.
To better visualize these properties, we plot the three bounds along with the HCRB in Fig.~\ref{f:dispsqueezing}(b) as a function of the squeezing parameter $r$ and for two different values of thermal photons $n$. 
We observe that for small values of thermal photons ($n=0.5$) and in the regime of small squeezing the HCRB is almost equal to the upper bound and slightly larger than the RLD-CRB, while for large squeezing the SLD-CRB becomes larger than the RLD-CRB and, as expected approaches the HCRB.
On the other hand, for larger values of thermal photons the gap between the RLD-CRB $C_{\theta }^{\mR}$ and the upper bound $\overline{C}_{\theta }^{\mathrm{H}}$ is so small that the HCRB is in fact almost indistinguishable from these bounds for all the values of squeezing considered; as expected, for large values of squeezing also the SLD-CRB converges to the HCRB.

\section{Discussion}
\label{sec:conclusions}

In this work, we have shown that the HCRB for Gaussian states can be efficiently computed as an SDP, even if the underlying Hilbert space is infinite-dimensional. 
This is possible because: (i) all constraints and figures of merit are expressed in terms of inner products between infinite-dimensional operators, and (ii) for Gaussian quantum statistical models one can restrict to a finite-dimensional basis of operators that are quadratic polynomials of canonical variables.
In this way, it is possible to obtain a compact description of the optimization problem in terms of finite-dimensional matrices.
Moreover, we have shown that this approach also allows us to understand SLD and RLD bounds for Gaussian states from a common perspective, and to show that they can also be computed as SDPs, similarly to the HCRB.

Although we have not pursued this aspect in this work, we note that the geometric approach based on inner products between operators naturally extends to the estimation of $q \leq p$ functions $\beta_j (\theta)$ of the original $p$ parameters.
In this case, the SLD-CRB and HCRB become minimizations over zero-mean operators, with slightly different local unbiasedness constraints $\Tr [ \partial_{\theta_k} \hat{\rho} \hat{X}_j  ] = \partial_{\theta_k} \beta_j $ (instead of $\delta_{jk}$ on the rhs), which do not change the SDP nature of the problem~\cite{Tsang2019}.
This approach can also accommodate singular quantum statistical models, with a non-invertible SLD-QFIM, as long as the functions $\beta_j(\theta)$ can be independently estimated (formally, if operators $\{ \hat{X}_j \}$ that satisfy these local unbiasedness constraints exist).
This is especially useful when $p=\infty$, i.e. for semiparametric estimation~\cite{Tsang2019}.
Such a framework is also suitable for analyzing precision limits for the estimation of functionals of the quantum state, assuming that the whole state is unknown.
This is closely related to classical shadow tomography~\cite{Huang2020}, which has recently been extended to CV systems~\cite{Gandhari2024,Becker2024}.

In recent years, new bounds for multiparameter quantum estimation have been proposed, aiming to capture more faithfully the behaviour of the fundamental bound $C^{\mathrm{MI}}_\theta(W)$, attainable when only single-copy measurements are available.
In this regard, a computable lower bound for finite-dimensional systems is the Nagaoka-Hayashi CRB~\cite{Conlon2020,Hayashi2023a}, which is also formulated as an SDP with stronger positive semidefinite constraints, expressed in terms of operators rather than inner products between them.
This makes the bound tighter than the HCRB; however, imposing such constraints on infinite-dimensional operators seems challenging, and a rigorous generalization of this bound for CV systems is still lacking, to the best of our knowledge.

Going beyond the local estimation paradigm adopted in this work, Bayesian versions of the Nagaoka-Hayashi CRB and of the HCRB have recently been introduced~\cite{Suzuki2024a,Zhang2024g}.
In particular, the techniques presented in this work are very likely applicable to evaluate the Bayesian HCRB of Ref.~\cite{Zhang2024g} for Gaussian states.
This is a first step to extend global estimation with Gaussian states~\cite{Morelli2020,Mukhopadhyay2024a} to the multiparameter domain.
More generally, studying non-asymptotic multiparameter estimation with CV systems is an interesting direction for further research~\cite{Meyer2023a}, similar questions have recently been tackled from the complementary point of view of learning theory in the context of state tomography~\cite{Mele2024a}.

Other interesting perspective for future studies in this area include: extending this framework to fermionic Gaussian systems, and studying multiparameter estimation with the further restrictions of using only Gaussian measurements readily available in optical platforms; results for this scenario are only known for specific single-parameter problems~\cite{Monras2006a,Oh2019a,Branford2019,Cenni2022}.

\section*{Acknowledgments}
F.~A. thanks Animesh Datta and Dominic Branford for fruitful discussions in the early stages of this project.
M.~G.~G. and F.~A. thank Matteo Paris and Alessio Serafini for several discussions on quantum estimation theory and Gaussian quantum states.
S.~C. acknowledges support by the China Scholarship Council.
M.~G.~G. acknowledges support from MUR and Next Generation EU via the NQSTI-Spoke2-BaC project QMORE (contract n. PE00000023-QMORE).
F.~A. acknowledges financial support from Marie Skłodowska-Curie Action EUHORIZON-MSCA-2021PF-01 (project \mbox{QECANM}, grant n. 101068347).

\bibliography{biblio-gaussian-hcrb}

\widetext
\section*{Appendices}
\appendix

In the appendices we extensively use Einstein’s summation convention, whereby repeated indexes in an expression are summed over the whole range of possible values, e.g. if $ j \in  [1,m]$, then $ f_j g_j = \sum_{j=1}^m  f_j g_j$.

\section{RLD inner product for a Gaussian state}
\label{a:traceproductformula}
Let us given a Gaussian quantum state $\rho_G$ characterized by a first moment vector $d$ and covariance matrix $\sigma$. One can prove the following formulas for the expectation values of product of canonical operators~\cite{Nichols2018}
\begin{align}
\text{Tr}\left[ \hat{\rho}_{G}\hat{r}_{l}\right]  &= d_{l},  \label{eq:product1} \\
\text{Tr}\left[ \hat{\rho}_{G}\hat{r}_{j}\hat{r}_{k}\right]  &= d_{j}d_{k}+%
\frac{1}{2}\left( \sigma _{jk}+i\Omega _{jk}\right) ,  \label{eq:product2} \\
\text{Tr}\left[\hat{\rho}_{G}\hat{r}_{j}\hat{r}_{k} \hat{r}_{m}\right]
&=d_{j}d_{k}d_{m}+\frac{1}{2}\left[ (\sigma _{jk}+i\Omega
_{jk})d_{m} \right.  + (\sigma _{km}+i\Omega _{km})d_{j} \left. +(\sigma _{jm}+i\Omega
_{jm})d_{k}\right] , \label{eq:product3} 
\end{align}
and
% \begin{widetext}
\begin{align}
\text{Tr}\left[ \hat{\rho}_{G}\hat{r}_{j}\hat{r}_{k}\hat{r}_{p}\hat{r}_{q}
\right]  
&= d_{j}d_{k}d_{p}d_{q}+\frac{1}{2}d_{p}d_{q}\sigma _{jk}+\frac{1}{2}%
d_{k}d_{q}\sigma _{jp}+\frac{1}{2}d_{j}d_{q}\sigma _{kp}  \notag \\
& \,\,\, +\frac{1}{2}d_{k}d_{p}\sigma _{jq}+\frac{1}{2}d_{j}d_{p}\sigma _{kq}+\frac{%
1}{2}d_{j}d_{k}\sigma _{pq}  \notag \\
& \,\,\, +\frac{i}{2}\left \{
\begin{array}{c}
\Omega _{jk}\left( d_{p}d_{q}+\frac{\sigma _{pq}}{2}\right) +\Omega
_{jp}\left( d_{k}d_{q}+\frac{\sigma _{kq}}{2}\right) +\Omega _{kp}\left(
d_{j}d_{q}+\frac{\sigma _{jq}}{2}\right)  \\
+\Omega _{jq}\left( d_{k}d_{p}+\frac{\sigma _{kp}}{2}\right) +\Omega
_{kq}\left( d_{j}d_{p}+\frac{\sigma _{jp}}{2}\right) +\Omega _{pq}\left(
d_{j}d_{k}+\frac{\sigma _{jk}}{2}\right)
\end{array}%
\right \}   \notag \\
&\,\,\, -\frac{1}{4}\left( \Omega _{jq}\Omega _{kp}+\Omega _{jp}\Omega
_{kq}+\Omega _{jk}\Omega _{pq}\right) +\frac{1}{4}\left( \sigma _{jq}\sigma
_{kp}+\sigma _{jp}\sigma _{kq}+\sigma _{jk}\sigma _{pq}\right) ,  \label{eq:product4}
\end{align}
% \end{widetext}

Let us consider now two operators that are at most linear and quadratic in the operators $\hat{r}$
\begin{align}
\hat{A} &= {a}^{(0)}\hat{\mathbb{I}} + ({a}^{(1)})^{\! \top} \hat{r} + \hat{r}^{\! \top}{a}^{(2)}\hat{r}\,, \\
\hat{B} &= {b}^{(0)}\hat{\mathbb{I}} + ({b}^{(1)})^{\! \top} \hat{r} + \hat{r}^{\! \top}{b}^{(2)}\hat{r}\,, 
%\label{eq:quadraticoperator}
\end{align}
where ${a}^{(0)}$ and ${b}^{(0)}$ are scalar complex numbers, ${a}^{(1)}, {b}^{(1)} \in \mathbb{C}^{2m}$, and where ${a}^{(2)}$ and ${b}^{(2)}$ are a $2m \times 2m$ complex matrices. Then one obtains
% \begin{widetext}
\begin{align}
    \text{tr}\left[ \hat{B}^{\dagger }\hat{\rho}_{G}\hat{A}\right]   & =\text{tr}\left[ \left( \left( b^{(0)}\right) ^{\ast }+\left( \emph{b}%
_{m}^{(1)}\right) ^{\ast }\hat{r}_{m}+\left( \emph{b}_{pq}^{(2)}\right)
^{\ast }\hat{r}_{p}\hat{r}_{q}\right) \hat{\rho}_{G}\left( a^{(0)}+\emph{a}%
_{l}^{(1)}\hat{r}_{l}+\emph{a}_{jk}^{(2)}\hat{r}_{j}\hat{r}_{k}\right) %
\right]   \notag \\
&=\left( b^{(0)}\right) ^{\ast }a^{(0)}+\left( b^{(0)}\right) ^{\ast }\emph{%
a}_{l}^{(1)}\text{tr}\left[ \hat{\rho}_{G}\hat{r}_{l}\right] +\left(
b^{(0)}\right) ^{\ast }\emph{a}_{jk}^{(2)}\text{tr}\left[ \hat{\rho}_{G}\hat{%
r}_{j}\hat{r}_{k}\right]   \notag \\
& +\left( \emph{b}_{m}^{(1)}\right) ^{\ast }a^{(0)}\text{tr}\left[ \hat{r}%
_{m}\hat{\rho}_{G}\right] +\left( \emph{b}_{m}^{(1)}\right) ^{\ast }\emph{a}%
_{l}^{(1)}\text{tr}\left[ \hat{r}_{m}\hat{\rho}_{G}\hat{r}_{l}\right]
+\left( \emph{b}_{m}^{(1)}\right) ^{\ast }\emph{a}_{jk}^{(2)}\text{tr}\left[
\hat{r}_{m}\hat{\rho}_{G}\hat{r}_{j}\hat{r}_{k}\right]   \notag \\
& +\left( \emph{b}_{pq}^{(2)}\right) ^{\ast }a^{(0)}\text{tr}\left[ \hat{r}%
_{p}\hat{r}_{q}\hat{\rho}_{G}\right] +\left( \emph{b}_{pq}^{(2)}\right)
^{\ast }\emph{a}_{l}^{(1)}\text{tr}\left[ \hat{r}_{p}\hat{r}_{q}\hat{\rho}%
_{G}\hat{r}_{l}\right]   \notag \\
& +\left( \emph{b}_{pq}^{(2)}\right) ^{\ast }\emph{a}_{jk}^{(2)}\text{tr}%
\left[ \hat{r}_{p}\hat{r}_{q}\hat{\rho}_{G}\hat{r}_{j}\hat{r}_{k}\right]
\end{align}
%\end{widetext}
By exploiting the relations in Eqs.~\eqref{eq:product1},~\eqref{eq:product2},~\eqref{eq:product3} and~\eqref{eq:product4}, one can write
%\begin{widetext}
\begin{align}
\Tr\left[ \hat{\rho}_{G}\hat{A}\hat{B}^{\dagger }\right]
&= \left( b^{(0)}\right) ^{\ast }a^{(0)}+\left( b^{(0)}\right) ^{\ast }\emph{%
a}_{l}^{(1)}d_{l}+\left( b^{(0)}\right) ^{\ast }\emph{a}_{jk}^{(2)}\left[
d_{j}d_{k}+\frac{1}{2}\left( \sigma _{jk}+i\Omega _{jk}\right) \right]
+\left( \emph{b}_{m}^{(1)}\right) ^{\ast }a^{(0)}d_{m}  \notag \\
& +\left( \emph{b}_{m}^{(1)}\right) ^{\ast }\emph{a}_{l}^{(1)}\left[
d_{l}d_{m}+\frac{1}{2}\left( \sigma _{lm}+i\Omega _{lm}\right) \right]
\notag \\
&+\left( \emph{b}_{m}^{(1)}\right) ^{\ast }\emph{a}_{jk}^{(2)}\left[
d_{k}d_{j}d_{m}+\frac{1}{2}\left[ (\sigma _{jk}+i\Omega _{jk})d_{m}+(\sigma
_{km}+i\Omega _{km})d_{j}+(\sigma _{jm}+i\Omega _{jm})d_{k}\right] \right]
\notag \\
&+\left( \emph{b}_{pq}^{(2)}\right) ^{\ast }a^{(0)}\left[ d_{p}d_{q}+\frac{1%
}{2}\left( \sigma _{pq}+i\Omega _{pq}\right) \right]   \notag \\
&+\left( \emph{b}_{pq}^{(2)}\right) ^{\ast }\emph{a}_{l}^{(1)}\left[
d_{p}d_{l}d_{q}+\frac{1}{2}\left[ (\sigma _{lp}+i\Omega _{lp})d_{q}+(\sigma
_{pq}+i\Omega _{pq})d_{l}+(\sigma _{lq}+i\Omega _{lq})d_{p}\right] \right]
\notag \\
& +\left( \emph{b}_{pq}^{(2)}\right) ^{\ast }\emph{a}_{jk}^{(2)}\left[
\begin{array}{c}
d_{j}d_{k}d_{p}d_{q}+\frac{1}{2}d_{p}d_{q}\sigma _{jk}+\frac{1}{2}%
d_{k}d_{q}\sigma _{jp}+\frac{1}{2}d_{j}d_{q}\sigma _{kp} \\
+\frac{1}{2}d_{k}d_{p}\sigma _{jq}+\frac{1}{2}d_{j}d_{p}\sigma _{kq}+\frac{1%
}{2}d_{j}d_{k}\sigma _{pq} \\
+\frac{i}{2}\left \{
\begin{array}{c}
\Omega _{jk}\left( d_{p}d_{q}+\frac{\sigma _{pq}}{2}\right) +\Omega
_{jp}\left( d_{k}d_{q}+\frac{\sigma _{kq}}{2}\right) +\Omega _{kp}\left(
d_{j}d_{q}+\frac{\sigma _{jq}}{2}\right)  \\
+\Omega _{jq}\left( d_{k}d_{p}+\frac{\sigma _{kp}}{2}\right) +\Omega
_{kq}\left( d_{j}d_{p}+\frac{\sigma _{jp}}{2}\right) +\Omega _{pq}\left(
d_{j}d_{k}+\frac{\sigma _{jk}}{2}\right)
\end{array}%
\right \}  \\
-\frac{1}{4}\left( \Omega _{jq}\Omega _{kp}+\Omega _{jp}\Omega _{kq}+\Omega
_{jk}\Omega _{pq}\right) +\frac{1}{4}\left( \sigma _{jq}\sigma _{kp}+\sigma
_{jp}\sigma _{kq}+\sigma _{jk}\sigma _{pq}\right)
\end{array}%
\right] \,,  \label{eq:productoperators_elements}
\end{align}%
% \end{widetext}
which can then be rewritten in a vectorial form as
% \begin{widetext}
\begin{align}
\Tr\left[ \hat{\rho}_{G}\hat{A}\hat{B}^{\dagger }\right]
&= \left( b^{(0)}\right) ^{\ast }a^{(0)}+\left( b^{(0)}\right) ^{\ast }d^{\! \top}%
\emph{a}^{(1)}+\left( b^{(0)}\right) ^{\ast }d^{\! \top}\emph{a}^{(2)}d+\frac{1}{2}%
\left( b^{(0)}\right) ^{\ast }\text{tr}\left[ \emph{a}^{(2)}\left( \sigma
-i\Omega \right) \right]   \notag \\
&+\left( \emph{b}^{(1)}\right) ^{\dagger }da^{(0)}+\left( \left( \emph{b}%
^{(1)}\right) ^{\dagger }d\right) \left( d^{\! \top}\emph{a}^{(1)}\right) +\frac{1%
}{2}\left( \emph{b}^{(1)}\right) ^{\dagger }\left( \sigma -i\Omega \right)
\emph{a}^{(1)}  \notag \\
&+\left( \left( \emph{b}^{(1)}\right) ^{\dagger }d\right) \left( d^{\! \top}\emph{%
a}^{(2)}d\right) +\frac{1}{2}\left( \left( \emph{b}^{(1)}\right) ^{\dagger
}d\right) \text{tr}\left[ \emph{a}^{(2)}\left( \sigma -i\Omega \right) %
\right] +\frac{1}{2}d^{\! \top}\emph{a}^{(2)}(\sigma +i\Omega )\left( \emph{b}%
^{(1)}\right) ^{\ast }  \notag \\
&+\frac{1}{2}\left( \emph{b}^{(1)}\right) ^{\dagger }(\sigma -i\Omega )%
\emph{a}^{(2)}d+a^{(0)}d^{\! \top}\left( \emph{b}^{(2)}\right) ^{\ast }d+\frac{1}{2%
}a^{(0)}\text{tr}\left[ \left( \emph{b}^{(2)}\right) ^{\ast }\left( \sigma
-i\Omega \right) \right]   \notag \\
&+\left( d^{\! \top}\left( \emph{b}^{(2)}\right) ^{\ast }d\right) \left( d^{\! \top}%
\emph{a}^{(1)}\right) +\frac{1}{2}\left( \emph{a}^{(1)}\right) ^{\! \top}(\sigma
+i\Omega )\left( \emph{b}^{(2)}\right) ^{\ast }d  \notag \\
&+\frac{1}{2}\left \{ \text{tr}\left[ \left( \emph{b}^{(2)}\right) ^{\ast
}\left( \sigma -i\Omega \right) \right] \right \} \left( d^{\! \top}\emph{a}%
^{(1)}\right) +\frac{1}{2}d^{\! \top}\left( \emph{b}^{(2)}\right) ^{\ast }\left(
\sigma -i\Omega \right) \emph{a}^{(1)}  \notag \\
&+\left( d^{\! \top}\left( \emph{b}^{(2)}\right) ^{\ast }d\right) \left( d^{\! \top}%
\emph{a}^{(2)}d\right) +\frac{1}{2}\left( d^{\! \top}\left( \emph{b}^{(2)}\right)
^{\ast }d\right) \text{tr}\left( \emph{a}^{(2)}\sigma \right)   \notag \\
&+\frac{1}{2}d^{\! \top}\left( \emph{b}^{(2)}\right) ^{\dagger }\sigma \emph{a}%
^{(2)}d+\frac{1}{2}d^{\! \top}\emph{a}^{(2)}\sigma \left( \emph{b}^{(2)}\right)
^{\ast }d+\frac{1}{2}d^{\! \top}\left( \emph{b}^{(2)}\right) ^{\ast }\sigma \emph{a%
}^{(2)}d  \notag \\
&+\frac{1}{2}d^{\! \top}\left( \emph{b}^{(2)}\right) ^{\ast }\sigma \left( \emph{a%
}^{(2)}\right) ^{\! \top}d+\frac{1}{2}\left[ \text{tr}\left( \left( \emph{b}%
^{(2)}\right) ^{\ast }\sigma \right) \right] \left( d^{\! \top}\emph{a}%
^{(2)}d\right)   \notag \\
&+\frac{i}{2}\left[ \left( d^{\! \top}\left( \emph{b}^{(2)}\right) ^{\ast
}d\right) \text{tr}\left[ \emph{a}^{(2)}\Omega ^{\! \top}\right] +\frac{1}{2}%
\left \{ \text{tr}\left[ \left( \emph{b}^{(2)}\right) ^{\ast }\sigma \right]
\right \} \left \{ \text{tr}\left[ \emph{a}^{(2)}\Omega ^{\! \top}\right] \right \} %
\right]   \notag \\
&+\frac{i}{2}\left[ d^{\! \top}\left( \emph{b}^{(2)}\right) ^{\dagger }\Omega ^{\! \top}%
\emph{a}^{(2)}d+\frac{1}{2}\text{tr}\left[ \emph{a}^{(2)}\sigma \left( \emph{%
b}^{(2)}\right) ^{\dagger }\Omega ^{\! \top}\right] \right]   \notag \\
&+\frac{i}{2}\left[ d^{\! \top}\emph{a}^{(2)}\Omega \left( \emph{b}^{(2)}\right)
^{\ast }d+\frac{1}{2}\text{tr}\left[ \left( \emph{b}^{(2)}\right) ^{\ast
}\sigma \emph{a}^{(2)}\Omega \right] \right]   \notag \\
&+\frac{i}{2}\left[ d^{\! \top}\left( \emph{b}^{(2)}\right) ^{\ast }\Omega ^{\! \top}%
\emph{a}^{(2)}d+\frac{1}{2}\text{tr}\left[ \left( \emph{b}^{(2)}\right)
^{\ast }\Omega ^{\! \top}\emph{a}^{(2)}\sigma \right] \right]   \notag \\
&+\frac{i}{2}\left[ d^{\! \top}\emph{a}^{(2)}\Omega \left( \emph{b}^{(2)}\right)
^{\dagger }d+\frac{1}{2}\text{tr}\left[ \left( \emph{b}^{(2)}\right)
^{\dagger }\sigma \emph{a}^{(2)}\Omega \right] \right]   \notag \\
&+\frac{i}{2}\left[ \left \{ \text{tr}\left[ \left( \emph{b}^{(2)}\right)
^{\ast }\Omega ^{\! \top}\right] \right \} \left( d^{\! \top}\emph{a}^{(2)}d\right) +%
\frac{1}{2}\left \{ \text{tr}\left[ \left( \emph{b}^{(2)}\right) ^{\ast
}\Omega ^{\! \top}\right] \right \} \left \{ \text{tr}\left[ \emph{a}^{(2)}\sigma %
\right] \right \} \right]   \notag \\
&+\frac{1}{4}\text{tr}\left[ \left( \emph{b}^{(2)}\right) ^{\ast }\Omega
\emph{a}^{(2)}\Omega \right] +\frac{1}{4}\text{tr}\left[ \left( \emph{b}%
^{(2)}\right) ^{\ast }\Omega \left( \emph{a}^{(2)}\right) ^{\! \top}\Omega \right]
+\frac{1}{4}\left \{ \text{tr}\left[ \left( \emph{b}^{(2)}\right) ^{\ast
}\Omega \right] \right \} \left \{ \text{tr}\left[ \emph{a}^{(2)}\Omega ^{\! \top}%
\right] \right \}   \notag \\
&+\frac{1}{4}\text{tr}\left[ \left( \emph{b}^{(2)}\right) ^{\ast }\sigma
\emph{a}^{(2)}\sigma \right] +\frac{1}{4}\text{tr}\left[ \left( \emph{b}%
^{(2)}\right) ^{\ast }\sigma \left( \emph{a}^{(2)}\right) ^{\! \top}\sigma \right]
+\frac{1}{4}\left \{ \text{tr}\left[ \left( \emph{b}^{(2)}\right) ^{\ast
}\sigma \right] \right \} \left \{ \text{tr}\left[ \emph{a}^{(2)}\sigma \right]
\right \} .  \label{eq:traceproductoperators}
\end{align}
% \end{widetext}
Notice that, given a Gaussian state with zero first moments ($d=0$) one gets
\begin{align}
\Tr\left[ \hat{\rho}_{G}\hat{A}\hat{B}^{\dagger }\right] 
& =  \left( b^{(0)}\right) ^{\ast }a^{(0)}+\frac{1}{2}\left( b^{(0)}\right)
^{\ast }\hbox{tr}\left[ \emph{a}^{(2)}\left( \sigma -i\Omega \right) \right] +\frac{%
1}{2}\left( \emph{b}^{(1)}\right) ^{\dagger }\left( \sigma -i\Omega \right)
\emph{a}^{(1)}  +\frac{1}{2}a^{(0)}\hbox{tr}\left[ \left( \emph{b}^{(2)}\right)
^{\ast }\left( \sigma -i\Omega \right) \right] \notag \\
& +\frac{i}{4}\hbox{tr}\left[ \left( \emph{b}^{(2)}\right) ^{\ast }\sigma \right] \hbox{tr}%
\left[ \emph{a}^{(2)}\Omega ^{\! \top}\right] +\frac{i}{4}\hbox{tr}\left[ \emph{a}%
^{(2)}\sigma \left( \emph{b}^{(2)}\right) ^{\dagger }\Omega ^{\! \top}\right] +%
\frac{i}{4}\hbox{tr}\left[ \left( \emph{b}^{(2)}\right) ^{\ast }\sigma \emph{a}%
^{(2)}\Omega \right]  \notag \\
&+\frac{i}{4}\hbox{tr}\left[ \left( \emph{b}^{(2)}\right) ^{\ast }\Omega ^{\! \top}\emph{%
a}^{(2)}\sigma \right] +\frac{i}{4}\hbox{tr}\left[ \left( \emph{b}^{(2)}\right)
^{\dagger }\sigma \emph{a}^{(2)}\Omega \right] +\frac{i}{4}\hbox{tr}\left[ \left(
\emph{b}^{(2)}\right) ^{\ast }\Omega ^{\! \top}\right] \hbox{tr}\left[ \emph{a}%
^{(2)}\sigma \right]  \notag \\
&+\frac{1}{4}\hbox{tr}\left[ \left( \emph{b}^{(2)}\right) ^{\ast }\Omega \emph{a}%
^{(2)}\Omega \right] +\frac{1}{4}\hbox{tr}\left[ \left( \emph{b}^{(2)}\right)
^{\ast }\Omega \left( \emph{a}^{(2)}\right) ^{\! \top}\Omega \right] +\frac{1}{4}\hbox{tr}%
\left[ \left( \emph{b}^{(2)}\right) ^{\ast }\Omega \right] \hbox{tr}\left[ \emph{a}%
^{(2)}\Omega ^{\! \top}\right]  \notag \\
&+\frac{1}{4}\hbox{tr}\left[ \left( \emph{b}^{(2)}\right) ^{\ast }\sigma \emph{a}%
^{(2)}\sigma \right] +\frac{1}{4}\hbox{tr}\left[ \left( \emph{b}^{(2)}\right)
^{\ast }\sigma \left( \emph{a}^{(2)}\right) ^{\! \top}\sigma \right] +\frac{1}{4}\hbox{tr}%
\left[ \left( \emph{b}^{(2)}\right) ^{\ast }\sigma \right] \hbox{tr}\left[ \emph{a}%
^{(2)}\sigma \right] ,  \label{18}
\end{align}%
% \end{widetext}
which can be compactly rewritten as Eq.~\eqref{eq:RLDquadratic} in terms of the block-diagonal matrix~\eqref{eq:Stheta}.

\section{Correspondence between phase space and Hilbert space superoperators}
\label{app:correspondence}

For operators acting on the right-hand side we have
\begin{eqnarray}
    \chi_{\hat{o}\hat{r}_{j}} (\tilde{r})  &=&
    %  &\leftrightarrow &
     \left[ -i\partial _{\tilde{r}_{j}}-%
    \frac{\Omega _{jk}}{2}\tilde{r}_{k}\right] \chi _{\hat{o}} (\tilde{r}) ,
    \label{eq:rightaction} \\
    \chi_{\hat{o}\hat{r}_{j}\hat{r}_{k}} (\tilde{r}) 
    %  &\leftrightarrow &
    &=& 
    \left[ -i\partial _{%
    \tilde{r}_{k}}- \frac{\Omega _{kl}}{2}\tilde{r}_{l}\right] \left[ -i\partial
    _{\tilde{r}_{j}} -  \frac{\Omega _{js}}{2}\tilde{r}_{s}\right] \chi _{\hat{o}} (\tilde{r})  \notag ,
\end{eqnarray}
while for operators acting on the left-hand side:
\begin{eqnarray}
    \chi_{\hat{r}_{j} \hat{o}} (\tilde{r}) &=&  \left[ - i \partial_{\tilde{r}_j} +  \frac{\Omega_{jk}}{2} \tilde{r}_k \right] \chi_{\hat{o}}(\tilde{r})  \\
    \chi_{ \hat{r}_{j}\hat{r}_{k} \hat{o} } (\tilde{r}) &=&     \left[ - i \partial_{\tilde{r}_j} +  \frac{\Omega_{jl}}{2} \tilde{r}_l \right]  \left[ - i \partial_{\tilde{r}_k} +  \frac{\Omega_{ks}}{2} \tilde{r}_s \right]  \chi_{\hat{o}}(\tilde{r}) 
\end{eqnarray}
From these we can find that the characteristic function of the commutator and anticommutator of a state with linear terms is~\cite[Eq.~(6.47)]{Serafini2023} 
\begin{equation}
    \chi_{\left[ \hat{r}_j , \hat{o} \right]} (\tilde{r}) 
     =  \Omega_{jk} \tilde{r}_k \chi_{\hat{o}} (\tilde{r})  , \qquad   \chi_{\left\{ \hat{r}_j , \hat{o} \right\}} (\tilde{r}) 
    =  - 2 i\partial_{\tilde{r}_j} \chi_{\hat{o}} (\tilde{r})  
    \label{eq:comm_anti_comm_chi_rj}
\end{equation}
and with quadratic terms:
\begin{equation}
    \label{eq:comm_anti_comm_chi_rkrj}
    \begin{aligned}
    \chi_{\left[ \hat{r}_k \hat{r}_j , \hat{o} \right]}
    %  \,\, \leftrightarrow  \, \, 
     = & - i\left(  \Omega_{j j'} \tilde{r}_{j'} \partial_{\tilde{r}_k} +  \Omega_{k k'} \tilde{r}_{k'}\partial_{\tilde{r}_j}  \right) \chi_{\hat{o}}  \\
     \chi_{\left\{ \hat{r}_k \hat{r}_j , \hat{o} \right\}}
    %   \,\, \leftrightarrow  \, \, 
    = & - 2 \partial_{\tilde{r}_k} \partial_{\tilde{r}_j} \chi_{\hat{o}} - i\Omega_{jk}\chi_{\hat{o}} + \frac{1}{2}  \Omega_{kk'} \Omega_{jj'} \tilde{r}_{k'} \tilde{r}_{j'}\chi_{\hat{o}} .
    \end{aligned}
\end{equation}

\section{Derivation of the RLD operators}
\label{a:RLDderivation}
In this section we will provide the derivation of the RLD-operator as in Eq.~(\ref{eq:RLDquadratic}), 
\begin{align}
    \hat{L}^{\mR} = l^{\mR(0)} \hat{\mathbb{I}} + (l^{\mR(1)})^{\! \top} \hat{r} + \hat{r}^{\! \top} l^{\mR(2)} \hat{r} \,, \label{eq:RLDstandardbasis_app}
\end{align}
that is in terms of the constant $l^{\mR(0)}$, the vector $l^{\mR(1)}$ and the matrix $l^{\mR(2)}$ reported in Eqs.~\eqref{eq:RLDstandardbasis0}, \eqref{eq:RLDstandardbasis1} and \eqref{eq:RLDstandardbasis2} (notice that we are here omitting the subscript that identifies one of the parameters to be estimated as it is completely irrelevant in this derivation). 

We start by reminding ourselves that the characteristic function of a Gaussian state $\rho_\theta$ characterized by a first moment vector $d$ and covariance matrix $\sigma$, can be written as~\cite{Serafini2023}
\begin{eqnarray}
\chi _{\hat{\rho}_{\theta }} &=&\text{Tr}\left[ \hat{D}_{-r}\hat{\rho}%
_{\theta }\right] =\exp \left( -\frac{1}{4}\sigma _{jk}\tilde{r}_{j}\tilde{r}_{k}+i\tilde{r}%
_{l}d_{l}\right) ,  \label{3}
\end{eqnarray}%
where $\hat{D}_{-r}$ is the displacement operator and $\tilde{r}=\Omega r$ with $r$=$\left( q_{1},p_{1},...,q_{n},p_{n}\right)^{\! \top}$ being a vector of 2$n$ real coordinates in phase-space. 

By differentiating the characteristic function $\chi _{\hat{\rho}_{\theta }}$ with respect to the parameter of
interest $\theta$, we have%
\begin{eqnarray}
\frac{\partial \chi _{\hat{\rho}_{\theta }}}{\partial \theta}
&=&\left( i\tilde{r}_{j}\frac{\partial d_{j}}{\partial \theta}-
\frac{1}{4}\frac{\partial \sigma _{jk}}{\partial \theta}\tilde{r}%
_{j}\tilde{r}_{k}\right) \chi _{\hat{\rho}_{\theta }}  \,.
\label{eq:derivative1}
\end{eqnarray}
Similarly, by recalling the definition of the RLD operator 
\begin{equation}
\frac{\partial\hat{\rho}_\theta}{\partial\theta} = \hat{\rho}_\theta \hat{L}^\mR \,,
\end{equation}
and by using Eq.~\eqref{eq:RLDstandardbasis_app} we can write the same derivative as
% \begin{widetext}
\begin{eqnarray}
\frac{\partial \chi _{\hat{\rho}_{\theta }}}{\partial \theta}
&=&\text{Tr}\left[ \hat{D}_{-r}\frac{\partial \hat{\rho}_{\theta }}{%
\partial \theta} \right]  \notag \\
&=&\text{Tr}\left[ \hat{D}_{-r}\hat{\rho}_{\theta }\left( l^{\mR(0)}+l_{ j}^{\mR(1)}\hat{r}_{j}+l_{jk}^{\mR(2)}\hat{r}_{j}\hat{r}_{k}\right) \right]  \notag \\
&=& l^{\mR(0)}\text{Tr}\left[ \hat{D}_{-r}\hat{\rho}_{\theta }\right]
+l_{j}^{\mR(1)}\text{Tr}\left[ \hat{D}_{-r}\hat{\rho}_{\theta }\hat{r}_{j}%
\right] +l_{jk}^{\mR(2)}\text{Tr}\left[ \hat{D}_{-r}\hat{\rho}_{\theta }%
\hat{r}_{j}\hat{r}_{k}\right]  \notag \\
&=&l^{\mR(0)}\chi _{\hat{\rho}_{\theta }}+l_{j}^{\mR(1)}\left[
-i\partial _{\tilde{r}_{j}}-\frac{\Omega _{jj^{\prime }}}{2}\tilde{r}%
_{j^{\prime }}\right] \chi _{\hat{\rho}_{\theta }}  
+l_{jk}^{\mR(2)}\left[ -i\partial _{\tilde{r}_{k}}-\frac{\Omega
_{kk^{\prime }}}{2}\tilde{r}_{k^{\prime }}\right] \left[ -i\partial _{\tilde{%
r}_{j}}-\frac{\Omega _{jj^{\prime }}}{2}\tilde{r}_{j^{\prime }}\right] \chi
_{\hat{\rho}_{\theta }}  \notag \\
&=&l^{\mR(0)}\chi_{\hat{\rho}_{\theta }}+l_{j}^{\mR(1)}\left[ d_{j}+%
\frac{1}{2}i\sigma_{jj^{\prime }}\tilde{r}_{j^{\prime }}-\frac{\Omega
_{jj^{\prime }}}{2}\tilde{r}_{j^{\prime }}\right] \chi _{\hat{\rho}_{\theta
}}  
+l_{jk}^{\mR(2)}\left[-(id_{k}-\frac{1}{2}\sigma _{kk^{\prime }}\tilde{r}%
_{k^{\prime }})(id_{j}-\frac{1}{2}\sigma _{jj^{\prime }}\tilde{r}_{j^{\prime
}})+\frac{1}{2}\sigma _{jk} \right.  \notag \\
&& \left.+\frac{1}{2}i\Omega _{jj^{\prime }}\left( \delta _{kj^{\prime }}+i\tilde{r}%
_{j^{\prime }}d_{k}-\frac{1}{2}\tilde{r}_{j^{\prime }}\sigma _{kk^{\prime }}%
\tilde{r}_{k^{\prime }}\right) +\frac{1}{2}i\Omega _{kk^{\prime }}\tilde{r}%
_{k^{\prime }}(id_{j}-\frac{1}{2}\sigma _{jj^{\prime }}\tilde{r}_{j^{\prime
}}) +\frac{1}{4}\Omega _{kk^{\prime }}\Omega _{jj^{\prime }}\tilde{r}%
_{k^{\prime }}\tilde{r}_{j^{\prime }}\right]\chi _{\hat{\rho}_{\theta }},
\label{eq:derivative2}
\end{eqnarray}%
% \end{widetext}
where we have used the identities in Eq.~\eqref{eq:rightaction}.
By equating the two formulas for the derivative \eqref{eq:derivative1} and \eqref{eq:derivative2}, we thus obtain 
% \begin{widetext}
\begin{eqnarray}
i\tilde{r}_{j}\frac{\partial d_{j}}{\partial \theta}-\frac{1}{4}%
\frac{\partial \sigma _{jk}}{\partial \theta}\tilde{r}_{j}\tilde{r}%
_{k} 
&=&l^{\mR(0)}+l_{j}^{\mR(1)}\left[ d_{j}+\frac{1}{2}i\sigma
_{jj^{\prime }}\tilde{r}_{j^{\prime }}-\frac{\Omega _{jj^{\prime }}}{2}%
\tilde{r}_{j^{\prime }}\right]  \notag \\
&&+l_{jk}^{\mR(2)}\left[-(id_{k}-\frac{1}{2}\sigma _{kk^{\prime }}\tilde{r}%
_{k^{\prime }})(id_{j}-\frac{1}{2}\sigma _{jj^{\prime }}\tilde{r}_{j^{\prime
}})+\frac{1}{2}\sigma _{jk} \right. \notag \\
&&+\frac{1}{2}i\Omega _{jj^{\prime }}\left( \delta _{kj^{\prime }}+i\tilde{r}%
_{j^{\prime }}d_{k}-\frac{1}{2}\tilde{r}_{j^{\prime }}\sigma _{kk^{\prime }}%
\tilde{r}_{k^{\prime }}\right) +\frac{1}{2}i\Omega _{kk^{\prime }}\tilde{r}%
_{k^{\prime }}(id_{j}-\frac{1}{2}\sigma _{jj^{\prime }}\tilde{r}_{j^{\prime
}})  \notag \\
&&\left.+\frac{1}{4}\Omega _{kk^{\prime }}\Omega _{jj^{\prime }}\tilde{r}%
_{k^{\prime }}\tilde{r}_{j^{\prime }}\right].  \label{eq:equalderivative}
\end{eqnarray}%
Now, we can start by considering the terms proportional to $\tilde{r}_j \tilde{r}_k$, that lead to the equation
\begin{eqnarray}
\frac{\partial \sigma _{jk}}{\partial \lambda _{\mu }} &=&l_{jk}^{\mR(2)}%
\left[ \sigma _{kk^{\prime }}\sigma _{jj^{\prime }}+i\Omega _{jj^{\prime
}}\sigma _{kk^{\prime }}+i\Omega _{kk^{\prime }}\sigma _{jj^{\prime
}}-\Omega _{kk^{\prime }}\Omega _{jj^{\prime }}\right]  \notag \\
&=&\sigma _{j^{\prime }j}l_{jk}^{\mR(2)}\sigma _{kk^{\prime }}-\left(
\Omega ^{\! \top}\right) _{j^{\prime }j}l_{jk}^{\mR(2)}\Omega _{kk^{\prime
}}+i\left( \Omega ^{\! \top}\right) _{j^{\prime }j}l_{jk}^{\mR(2)}\sigma
_{kk^{\prime }}+i\sigma _{j^{\prime }j}l_{jk}^{\mR(2)}\Omega _{kk^{\prime
}}  \notag \\
&=&\sigma _{j^{\prime }j}l_{jk}^{\mR(2)}\sigma _{kk^{\prime }}+\Omega
_{j^{\prime }j}l_{jk}^{\mR(2)}\Omega _{kk^{\prime }}-i\Omega _{j^{\prime
}j}l_{jk}^{\mR(2)}\sigma _{kk^{\prime }}+i\sigma _{j^{\prime }j}l_{jk}^{\mR(2)}\Omega _{kk^{\prime }}  \,,
\label{12}
\end{eqnarray}%
% \end{widetext}
which in terms of matrices can be written as
\begin{equation}
\frac{\partial \sigma }{\partial \theta}  = \sigma l^{\mR(2)}\sigma +\Omega l^{\mR(2)}\Omega -i\Omega l^{\mR(2)}\sigma
+i\sigma l^{\mR(2)}\Omega =\left( \sigma -i\Omega \right) l^{\mR(2)}\left( \sigma +i\Omega
\right) \,.
%&=&\left( \sigma +i\Omega \right) l^{\mR(2)\dagger }\left( \sigma -i\Omega \right) ,  \label{13}
\end{equation}
This lead to the following formula for the matrix $l^{\mR(2)}$
\begin{align}
l^{\mR(2)} = \left( \sigma -i\Omega \right)^{-1} ( \partial_\theta \sigma) \left( \sigma +i\Omega
\right)^{-1}\,,
\end{align}
which, by means of Eq.~(\ref{eq:vectorproduct}), can be written in terms of the vectorized matrices as
\begin{align}
    \text{vec}\left[ l^{\mR(2)}\right] &= \left[ \left( \sigma -i\Omega
\right) \otimes \left( \sigma -i\Omega \right) \right] ^{-1}\text{vec}\left[\partial_\theta \sigma \right] \,.
\end{align}
Similarly we can now consider the terms linear in $\tilde{r}_j$ in Eq.~\eqref{eq:equalderivative}, leading to the equation
% \begin{widetext}
\begin{eqnarray}
\frac{\partial d_{j}}{\partial \theta} &=&l_{j}^{\mR(1)}\left[
\frac{1}{2}\sigma _{jj^{\prime }}+\frac{1}{2}i\Omega _{jj^{\prime }}\right]
+l_{jk}^{\mR(2)}\left[ \frac{1}{2}d_{k}\sigma _{jj^{\prime }}+\frac{1}{2}%
d_{j}\sigma _{kk^{\prime }}+\frac{1}{2}i\Omega _{jj^{\prime }}d_{k}+\frac{1}{%
2}i\Omega _{kk^{\prime }}d_{j}\right]  \notag \\
&=&\frac{1}{2}l_{j}^{\mR(1)}\left[ \sigma _{jj^{\prime }}+i\Omega
_{jj^{\prime }}\right] +\frac{1}{2}l_{jk}^{\mR(2)}\left[ d_{k}\sigma
_{jj^{\prime }}+d_{j}\sigma _{kk^{\prime }}+i\Omega _{jj^{\prime
}}d_{k}+i\Omega _{kk^{\prime }}d_{j}\right]  \notag \\
&=&\frac{1}{2}l_{j}^{\mR(1)}\left[ \sigma _{jj^{\prime }}+i\Omega
_{jj^{\prime }}\right] +\frac{1}{2}\left[ \sigma _{jj^{\prime }}+i\Omega
_{jj^{\prime }}\right] l_{jk}^{\mR(2)}d_{k}+\frac{1}{2}\left[ \sigma
_{kk^{\prime }}+i\Omega _{kk^{\prime }}\right] l_{jk}^{\mR(2)}d_{j}  \notag
\end{eqnarray}
which in vectorial form corresponds to
\begin{eqnarray}
\frac{\partial d}{\partial \theta}
&=&\frac{1}{2}(\sigma -i\Omega )l^{\mR(1)}+\frac{1}{2}(\sigma -i\Omega
)l^{\mR(2)}d+\frac{1}{2}(\sigma -i\Omega )\left(l^{\mR(2)}\right)
^{\! \top}d.  \label{10}
\end{eqnarray}
% \end{widetext}
and thus lead to the following equation for the vector $l^{\mR(1)}$
\begin{eqnarray}
l^{\mR(1)} &=&2(\sigma -i\Omega )^{-1} (\partial_\theta d)  - \left( l^{\mR(2)} +  l^{\mR(2) \top} \right) d .
\label{11}
\end{eqnarray}
Finally, by considering the constant terms in Eq.~\eqref{eq:equalderivative}, we obtain
\begin{equation}
l^{\mR(0)}+l_{j}^{\mR(1)}d_{j}+l_{jk}^{\mR(2)}(d_{j}d_{k}+\frac{1}{2%
}\sigma _{jk}+\frac{1}{2}i\Omega _{jk})=0\,,
\end{equation}
leading to the following equation for the constant term $l^{\mR(0)}$
\begin{eqnarray}
l^{\mR(0)} &=&-\frac{1}{2}\text{Tr}\left[ (\sigma -i\Omega )l^{\mR(2)}\right] -l^{\mR(1)^{\! \top}}d-d^{\! \top}l^{\mR(2)}d\,. \notag \\
\end{eqnarray}

\section{Explicit form of Eq.~\eqref{eq:Dsuper_chi} for Gaussian states and quadratic operators}
\label{app:explicit_Dsuper}

For completeness, we present an explicit derivation which is slightly simplified by the use of the central basis.
We start by writing Eq.~\eqref{eq:Dsuper_chi} for quadratic $\hat{X}$ and $\hat{Z}$, using the explicit mapping of the action of canonical operators on the characteristic function as differential operators presented in Appendix~\ref{app:correspondence}.
Without loss of generality, we assume $\hat{X}$ and $\hat{Z}$ to be zero-mean. 
Even when $\hat{X}$ is not zero-mean, $\hat{Z}$ is zero-mean, since the trace of the commutator on the right-hand side of Eq.~\eqref{eq:Dsuper_def} is zero.
Therefore any D-invariant subspace must be a subspace of zero-mean operators.
We can thus express these operators in the central basis, as in Eq.~\eqref{eq:centralbasis}.

It is useful to notice that we the characteristic function of an arbitrary quadratic operators $\hat{A}$ acting from left or from right on a Gaussian state $\hat{\rho}_\theta$, has a simple relationship to the characteristic function of its displaced version $\hat{X}=\hat{D}_{d_{\theta}} \hat{A} \hat{D}_{d_\theta}^\dag$ applied to the same state with the first moments set to zero $\hat{\rho}_\theta^0$ (these are the same objects appearing in Eq.~\eqref{eq:TrAB_corrisp_TrXY}).
One can prove by direct calculation that
\begin{align}
    \label{eq:anticomm_chi_zeromean}
    \chi_{\hat{A}\hat{\rho}_{\theta} + \hat{\rho}_\theta \hat{A} } ( r ) =  \chi_{ \hat{X}\hat{\rho}_\theta^0 + \hat{\rho}_\theta^{0} \hat{X} } ( r ) e^{i d_\theta ^\top \Omega r } \\ 
    \chi_{\hat{A}\hat{\rho}_{\theta} - \hat{\rho}_\theta \hat{A} } ( r ) =  \chi_{ \hat{X}\hat{\rho}_\theta^0 - \hat{\rho}_\theta^{0} \hat{X} } ( r ) e^{i d_\theta ^\top \Omega r } \,, \label{eq:comm_chi_zeromean}
\end{align}
which relates the characteristic functions of commutators and anticommutators of zero-mean operators with a Gaussian state $ \hat{\rho}_\theta$ with first moments $d_\theta$, to those obtained from a state $\hat{\rho}_\theta^0$ with zero first moments.

Thus, the characteristic functions in Eq.~\eqref{eq:anticomm_chi_zeromean} and~\eqref{eq:comm_chi_zeromean} have the same phase factor, which can be simplified in Eq.~\eqref{eq:Dsuper_chi} and we can set the first moments of the Gaussian state to zero without loss of generality.
Now we can exploit the general phase-space relations reported in Appendix~\ref{app:correspondence}, Eqs.~\eqref{eq:comm_anti_comm_chi_rj} and~\eqref{eq:comm_anti_comm_chi_rkrj}, which we explicitly evaluate here for a Gaussian state with zero first moments, obtaining the linear terms
\begin{align}
    \chi_{\left[ \hat{r}_j , \hat{\rho}_G \right]}(\tilde{r}) = & \sum_{k} \Omega_{jk} \tilde{r}_k \chi_{\hat{\rho}_{G}}(\tilde{r})\\ 
    \chi_{\left\{ \hat{r}_j , \hat{\rho}_G \right\}}(\tilde{r}) = &   i \sum_{j'} \sigma_{j j'} \tilde{r}_{j'} \chi_{\hat{\rho}_{G}}(\tilde{r})  \, ,
\end{align}
and the quadratic ones
\begin{align}
    \chi_{\left[ \hat{r}_j \hat{r}_k , \hat{\rho}_G \right]} (\tilde{r})    = &  \frac{\I}{2} \sum_{lm} \left( \Omega_{kl} \sigma_{jm} + \Omega_{jl} \sigma_{km} \right) \tilde{r}_l \tilde{r}_m \chi_{\hat{\rho}_{G}}(\tilde{r}) \\ 
    \chi_{\left\{ \hat{r}_j \hat{r}_k , \hat{\rho}_G \right\}}(\tilde{r}) =
      & \sum_{lm} 
     \left( - \frac{1}{2} \sigma_{jl} \sigma_{k m}   + \frac{1}{2} \Omega_{k l}\Omega_{jm} \right)\tilde{r}_l \tilde{r}_m \chi_{\hat{\rho}_{G}}(\tilde{r}) + \left( \sigma_{jk} -i\Omega_{jk} \right)\chi_{\hat{\rho}_{G}} (\tilde{r}) .
\end{align}
Thus, we can simplify the characteristic function $\chi_{\hat{\rho}_{G}}$ on the two sides of Eq.~\eqref{eq:Dsuper_chi}, since it is never zero, and equating the coefficients for the various terms of the quadratic polynomial we obtain the following equations (in vectorized form)
\begin{align}
    \sigma Z^{(1)}  &= \Omega X^{(1)}  \\ 
    \left( \sigma \otimes \sigma - \Omega \otimes \Omega \right) \mathrm{vec}\left[ Z^{(2)} \right] &= \left( \Omega \otimes \sigma + \sigma \otimes \Omega \right) \mathrm{vec}\left[ X^{(2)} \right] \, , \nonumber
\end{align}
equivalent to the more compact notation
\begin{equation}
    \Re( S_\theta ) \bar{Z} = -\Im( S_\theta) \bar{X},
\end{equation}
in terms of the matrix~\eqref{eq:Stheta} and vectors \eqref{eq:barvector}.
We have omitted the equations for the terms proportional to the characteristic function, since they are always satisfied by the assumption that $\hat{Z}$ is zero-mean.
Evidently, solving this equation is analogous to solving the Lyapunov equation for the SLD operator and it can be done simply by inverting $\Re (S_\theta)$.
This matrix is invertible for states with no pure normal modes, i.e. all symplectic eigenvalues strictly greater than one~\cite{Serafini2023}, see also the remarks at the end of Sec.~\ref{sec:derivation} for the general case.

\section{Additional results for joint estimation of phase and loss}
\label{a:appendix_phaseloss}
We here report some additional analytical results for the joint estimation of phase and loss via either coherent states ($r=0$) or squeezed vacuum states ($\alpha=0$) as input, but for generic values of the loss parameter $\eta$. 
We start by reporting the general formula for the commutator between the two SLD operators for input coherent states, that reads
\begin{equation}
\left[ \hat{L}_{\varphi }^{\mS},\hat{L}_{\eta }^{\mS}\right] =i\left[ \frac{%
4\eta \left\vert \alpha \right\vert ^{2}}{1-\eta }+2\Gamma _{1}\hat{x}%
_{1}+2\Gamma _{2}\hat{p}_{1}\right] , 
\nonumber
\end{equation}%
where%
\begin{eqnarray}
\Gamma _{1} &=&\frac{\sqrt{2}(1-2\eta )(\Re(\alpha )\cos \varphi +%
\Im(\alpha )\sin \varphi )}{\sqrt{\eta }(1-\eta )},  \notag \\
\Gamma _{2} &=&\frac{\sqrt{2}(1-2\eta )(\Im(\alpha )\cos \varphi -%
\Re(\alpha )\sin \varphi )}{\sqrt{\eta }(1-\eta )}. 
\nonumber
\end{eqnarray}

For an input squeezed vacuum state ($\alpha =0$) we obtain
\begin{equation}
\left[ \hat{L}_{\varphi }^{\mS},\hat{L}_{\eta }^{\mS}\right] =4i\left[ \left(
\Upsilon _{1}\hat{x}_{1}+\Upsilon _{3}\hat{p}_{1}\right) \hat{x}_{1}+\left(
\Upsilon _{3}\hat{x}_{1}+\Upsilon _{2}\hat{p}_{1}\right) \hat{p}_{1}\right] ,
\nonumber
\end{equation}%
where%
\begin{eqnarray}
\Upsilon _{1} &=&\frac{(1-e^{-4r})\left[ -\eta ^{2}\cos ^{2}\varphi +(1-\eta
)^{2}e^{2r}\cos 2\varphi +\eta ^{2}e^{4r}\sin ^{2}\varphi \right] }{4(1-\eta
)\left[ 1-\eta (1-\eta )+\eta (1-\eta )\cosh 2r\right] ^{2}},  \notag \\
\Upsilon _{2} &=&\frac{(1-e^{-4r})\left[ \eta ^{2}e^{4r}\cos ^{2}\varphi
-(1-\eta )^{2}e^{2r}\cos 2\varphi -\eta ^{2}\sin ^{2}\varphi \right] }{%
4(1-\eta )\left[ 1-\eta (1-\eta )+\eta (1-\eta )\cosh 2r\right] ^{2}},
\notag \\
\Upsilon _{3} &=&\frac{\left[ \eta ^{2}\cosh 2r-(1-\eta )^{2}\right] \sin
2\varphi \sinh 2r}{2(1-\eta )\left[ 1-\eta (1-\eta )+\eta (1-\eta )\cosh 2r%
\right] ^{2}}.  \nonumber
\end{eqnarray}%

As regards the Uhlmann curvature matrix, for inut coherent states we find
\begin{equation}
\mathcal{I}=
\begin{pmatrix}
0 & 2\left\vert \alpha \right\vert ^{2} \\
-2\left\vert \alpha \right\vert ^{2} & 0%
\end{pmatrix},  \nonumber
\end{equation}%
which is independent of $\eta$, while for the input squeezed vacuum states, we have

\begin{equation}
\mathcal{I}=
\begin{pmatrix}
0 & \frac{\eta \sinh ^{2}2r}{\left[ 1-\eta (1-\eta )+\eta (1-\eta )\cosh 2r%
\right] ^{2}} \\
-\frac{\eta \sinh ^{2}2r}{\left[ 1-\eta (1-\eta )+\eta (1-\eta )\cosh 2r%
\right] ^{2}} & 0%
\end{pmatrix} .  \nonumber
\end{equation}

\begin{figure}
\centering 
\includegraphics[width=0.6\columnwidth]{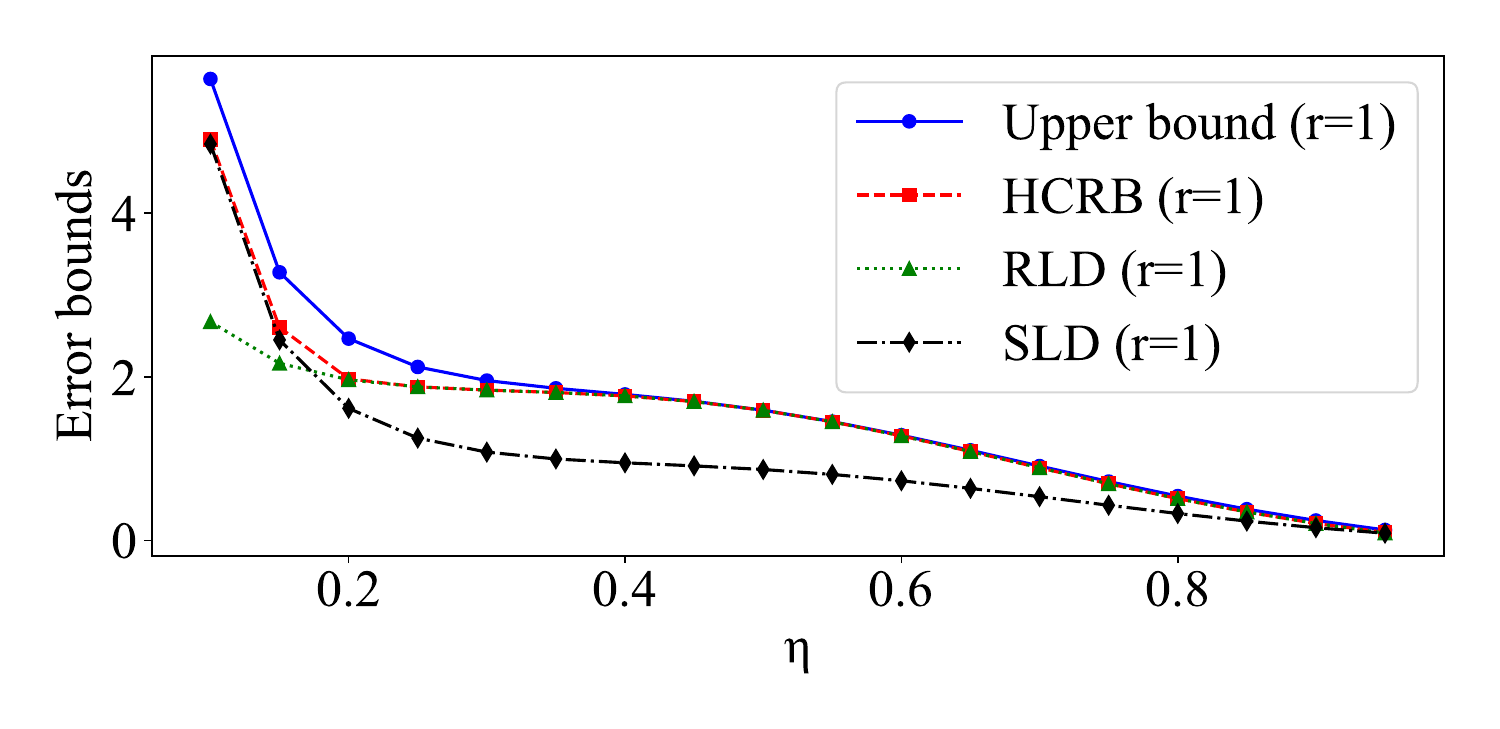} 
\caption{{}(Color online) Error bounds\textbf{\ }as a function of the loss parameter
$\eta$ for an input squeezed vacuum state ($\alpha=0$ and $r=1.0$).
\label{f:phaseloss_squeezed_r1}}
\end{figure}

Finally, we can report the analytical formulas of SLD-CRB $%
C_{(\varphi ,\eta )}^{\mS}$, RLD-CRB $C_{(\varphi ,\eta )}^{\mR}$, and of the
upper bound to the HCRB $\overline{C}_{(\varphi ,\eta )}^{\mathrm{H}}$, for input coherent states read

\begin{eqnarray}
C_{(\varphi ,\eta )}^{\mS} &=&\frac{1+4\eta ^{2}}{4\eta \left\vert \alpha
\right\vert ^{2}},  \notag \\
\overline{C}_{(\varphi ,\eta )}^{\mathrm{H}} &=&C_{(\varphi ,\eta )}^{\mR}=C_{(\varphi ,\eta
)}^{\mS}+\frac{1}{\left\vert \alpha \right\vert ^{2}},  \nonumber
\end{eqnarray}
and for input squeezed vacuum states, 
% \begin{widetext}
\begin{eqnarray}
C_{(\varphi ,\eta )}^{\mS} &=&\frac{16\eta ^{4}(1-\eta )^{2}+\left[ 1+2\eta
(\eta +4\eta ^{2}-4\eta ^{3}-1)\right] \text{csch}^{2}r-\left[ 1-2\eta
(1-\eta )\right] ^{2}\text{sech}^{2}r}{8\eta ^{2}\left[ 1-2\eta (1-\eta )%
\right] },  \notag \\
C_{(\varphi ,\eta )}^{\mR} &=&C_{(\varphi ,\eta )}^{\mS}+\frac{\left[ 4\eta
(1+\eta -2\eta ^{2})-1\right] \text{csch}^{2}r+(1-2\eta )^{2}\text{sech}^{2}r%
}{8\eta ^{2}\left[ 1-2\eta (1-\eta )\right] },  \notag \\
\overline{C}_{(\varphi ,\eta )}^{\mathrm{H}} &=&C_{(\varphi ,\eta )}^{\mS}+\frac{(1-\eta )\text{%
csch}^{2}r}{1-2\eta (1-\eta )}=C_{(\varphi ,\eta )}^{\mR}+\frac{(1-2\eta )^{2}%
\text{csch}^{2}2r}{2\eta ^{2}\left[ 1-2\eta (1-\eta )\right] }.  \nonumber
\end{eqnarray}%
% \end{widetext}
From the formulas above, we thus observe how the upper bound $\overline{C}_{(\varphi ,\eta )}^{\mathrm{H}}$ coincides the RLD-CRB $C_{(\varphi ,\eta )}^{\mR}$ and thus with the HCRB $C_{(\varphi ,\eta )}^{\mathrm{H}}$, independently on the values of $\eta$. Differently, for input squeezed vacuum states one has the $\overline{C}_{(\varphi ,\eta )}^{\mathrm{H}}=C_{(\varphi ,\eta )}^{\mR}$ only for $\eta=1/2$, showing that in general the HCRB does not coincide with the RLD bound. This is confirmed by the numerical results for the HCRB shown in Fig.~\ref{f:phaseloss_squeezed_r1}, where we observe how for an input squeezed vacuum state with $r=1.0$ the HCRB $C_{(\varphi ,\eta )}^{\mathrm{H}}$ is in general larger than the RLD-CRB $C_{(\varphi ,\eta )}^{\mR}$, apart from the region $\eta \gtrsim 0.3$, where the two bounds almost coincide.

\end{document}